\documentclass[acmtog]{acmart} %
\acmSubmissionID{1260}

\usepackage{booktabs} %
\usepackage{framed}
\usepackage{graphicx,subcaption}
\usepackage{svg}
\usepackage{bbm}

\usepackage[capitalize]{cleveref}
\crefname{algocf}{Alg.}{Algs.}
\Crefname{algocf}{Algorithm}{Algorithms}
\crefname{section}{Sec.}{Secs.}
\Crefname{section}{Sec.}{Secs.} %

\usepackage{csvsimple}

\usepackage{siunitx}
\usepackage{float}

\usepackage[ruled]{algorithm2e} %

\SetAlFnt{\small}
\SetAlCapFnt{\small}
\SetAlCapNameFnt{\small}
\SetAlCapHSkip{0pt}

\def\tp{^{\mathsf{T}}}
\def\Bz/{B{\'{e}}zier}

\newcommand{\R}{{\mathbb R}}
\newcommand{\mv}[1]{\mathbf{#1}}

\DeclareMathOperator*{\argmin}{arg\,min}

\newcommand{\dfdx}[2]{\ensuremath \partial_{#2} #1}
\newcommand{\dfdxfrac}[2]{\ensuremath \frac{\partial #1}{\partial #2}}

\newcommand{\shape}{\Omega}
\newcommand{\surface}{\partial \Omega}
\newcommand{\normal}{\mv{n}}
\newcommand{\point}{\mv{x}}
\newcommand{\indicator}{\chi_{\shape}}
\newcommand{\winding}{w}
\newcommand{\windingsub}{\winding_{\surface}}
\newcommand{\windingsmooth}{\tilde{\winding}}
\newcommand{\kernel}{k}
\newcommand{\sceneparam}{\theta}
\newcommand{\params}{\boldsymbol{\theta}}
\newcommand{\region}{D}

\newcommand{\loss}{E}

\newcommand{\mymathbox}[3]{%
{%
\setlength{\fboxrule}{1pt}%
\setlength{\fboxsep}{3pt}%
\sbox0{$\displaystyle #3$}%
\mathord{%
\raisebox{0.45\baselineskip}{\hbox{%
\setlength{\tabcolsep}{0pt}%
\renewcommand{\arraystretch}{0.9}%
\begin{tabular}{c}
\fcolorbox{#1}{white}{\footnotesize\bfseries #2}\\[-0.6ex]
\fcolorbox{#1}{white}{\usebox0}
\end{tabular}%
}}%
}%
}%
}

\SetKwComment{Comment}{$\triangleright\;$}{}

\usepackage{tikz}

\usetikzlibrary{backgrounds}

\makeatletter
 \newcommand{\linkdest}[1]{\Hy@raisedlink{\hypertarget{#1}{}}}
\makeatother

\newcommand{\define}[1]{\linkdest{#1}{\emph{#1}}}

\setcopyright{rightsretained}

\copyrightyear{2026}
\acmYear{2026}
\setcopyright{cc}
\setcctype{by}
\acmConference[SIGGRAPH Conference Papers '26]{Special Interest Group on Computer Graphics and Interactive Techniques Conference Conference Papers}{July 19--23, 2026}{Los Angeles, CA, USA}
\acmBooktitle{Special Interest Group on Computer Graphics and Interactive Techniques Conference Conference Papers (SIGGRAPH Conference Papers '26), July 19--23, 2026, Los Angeles, CA, USA}
\acmDOI{10.1145/3799902.3811203}
\acmISBN{979-8-4007-2554-8/2026/07}

\begin{document}
\title{Differentiable Voxelization of Surface Representations}

\author{Tobias Djuren}
\orcid{0009-0003-7043-9983}
\affiliation{%
 \institution{Technische Universität Berlin}
 \city{Berlin}
 \country{Germany}}
\email{t.djuren@tu-berlin.de}

\author{Ugo Finnendahl}
\orcid{0000-0002-7098-1524}
\authornote{Equal contribution (shared second author).}
\affiliation{%
 \institution{Technische Universität Berlin}
 \city{Berlin}
 \country{Germany}
}
\email{finnendahl@tu-berlin.de}

\author{Markus Worchel}
\orcid{0000-0002-3469-6750}
\authornotemark[1]
\affiliation{%
 \institution{Technische Universität Berlin}
 \city{Berlin}
 \country{Germany}
}
\email{m.worchel@tu-berlin.de}

\author{Hendrik Meyer}
\orcid{0009-0002-4803-750X}
\affiliation{%
 \institution{Technische Universität Berlin}
 \city{Berlin}
 \country{Germany}
}
\email{hendrik.meyer@tu-berlin.de}

\author{Marc Alexa}
\orcid{0000-0002-9854-8466}
\affiliation{%
 \institution{Technische Universität Berlin}
 \city{Berlin}
 \country{Germany}
}
\email{marc.alexa@tu-berlin.de}

\renewcommand\shortauthors{Djuren, Finnendahl, Worchel, Meyer, and Alexa}

\begin{teaserfigure}
\includegraphics[width=\columnwidth]{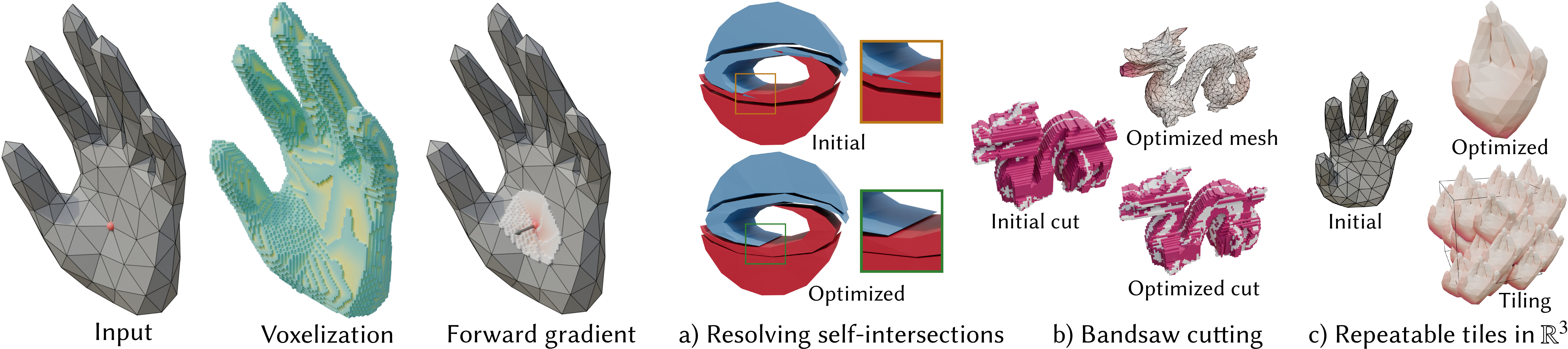}
    \centering
    \caption{We present an efficient technique for differentiating voxel values with respect to the surface that generates the voxel grid. 
    This enables the solution of various problems that can easily be solved on regular voxel grids, such as intersection detection, for surface representations like triangle meshes.
    The technique can be readily integrated into gradient-based optimization frameworks like PyTorch~\cite{Paszke2019}. Moving the highlighted vertex (left) in the indicated direction changes the voxelization of all adjacent triangles (right). 
    We discuss various applications, like a) the resolving of self-intersections,  b) optimized cuts for manufacturing with bandsaws, and c) shape deformation to form repeatable tilings that fill the $\R^3$.
    }
    \label{fig:teaser}
\end{teaserfigure}

\begin{abstract}
Different shape representations facilitate different computations. Surface representations, in particular meshes, are often used for modeling, whereas volume representations are useful for spatial queries such as intersection or containment. 
Optimizing a surface representation based on a volumetric properties by gradient descent requires the derivatives of the volume relative to its bounding surface. We derive this gradient for winding numbers and show that it can be efficiently computed for volumetric values sampled on a regular grid (voxel representation) and surface parameters based on vertex sets (triangle meshes). This enables an efficient solution for a variety of optimization problems. We demonstrate the practical use of this approach at the examples of deforming meshes to resolve intersections, being manufacturable by cutting with a bandsaw from three directions, and creating shapes that are close to tiling 3D space. 
\end{abstract}


\maketitle

\section{Introduction}

The shape of a solid object can be represented in two fundamentally different ways: (1) as the part of space $\shape \subset \R^n$ covered by the object; or as the boundary $\surface$ enclosing its interior. We will refer to these representations as \emph{volume} and \emph{surface} representations. Each class comprises various instances, serving different purposes and application domains. 

In many applications, the interior of the shape is not visible or irrelevant. Then surface representations suffice to describe the shape and avoid storing information that is irrelevant. 
The most relevant surface representations are polygon meshes~\cite{botsch2010polygon}, dominating in the visual effects and games industry, parametric surfaces used in CAGD~\cite{farin2002curves}, and subdivision surfaces~\cite{warren2001subdivision} being a combination of the two. 

Volumetric representations are useful when either the interior is heterogeneous or the space covered by the object, possibly also in relation to other shapes, is relevant. Examples are intersection queries~\cite{10.2312:egtp.19871000} or constructive solid geometry~\cite{1646334}. Clearly, the most relevant way to store information about the interior is a regular grid, so-called \emph{voxel} representations. Voxel grids align well with hardware and, in particular, memory access, making algorithms operating on them very efficient in practice. This is likely part of the reason why they are used in compute-heavy domains, for example, in machine learning~\cite{NEURIPS2020_b4b75896,Sun_2022_CVPR,Deng_Shi_Li_Zhou_Zhang_Li_2021}. The information stored in the voxels varies: binary information about being interior (occupancy), distance to the closest point on the surface (distance fields), or, as we suggest here, the winding number~\cite{mobius1865bestimmung} -- see \cref{sec:differentiable_voxelization} for details.

Modeling and optimization of shapes relies almost exclusively on defining an \emph{energy function} $E$ (similar to the \emph{loss} used in machine learning) that takes on smaller values for better solutions to the problem. Our original motivation for this work was modeling geometry that can be cut out of a solid using a bandsaw. This requires the object to be the intersection of extrusions (see Sec.~\ref{sec:bandsaw} for details). 

While this condition can be easily checked for a volumetric representation, it is quite cumbersome for surfaces. 
Yet for exploring the space of deformation of an object, surface representations have shown to be very useful~\cite{Botsch:deformation,MONTAGNAT20011023}. 
We believe this a common scenario: a shape is given in a surface representation, governed by parameters $\params \in \R^m$ (e.g. the vertex positions of a polygon mesh); we wish to find or approximate $\argmin_{\params} E(\shape(\params))$, where the energy $E$ depends on the volumetric representation $\shape(\params)$ of the shape.
Efficient optimization relies on the gradient of the energy $E$ relative to the parameters $\params$, which by the chain rule is a product of two quantities:
\begin{equation}
    \frac{\partial E(\shape(\params))}{\partial \params} = \frac{\partial E(\shape(\params))}{\partial\; \shape} \cdot \frac{\partial\; \shape}{\partial \params}
\end{equation}
While the derivative(s) of the energy with respect to the volume representation will depend on the energy, the derivatives of the volume with respect to the parameters of the surface representation are a general problem, whose solution enables a range of methods and applications. 
In \cref{sec:differentiable_voxelization} we derive a general continuous solution for the winding number function, allowing the treatment of self-intersecting surfaces. 
For the most common case of triangle meshes, we show that this approach allows computing the derivatives for a voxel grid in a highly efficient way (\cref{sec:closed_form_voxelization}). 

We evaluate the implementation of this approach (\cref{sec:implementation}) and then demonstrate its effectiveness and versatility in several applications: apart from the mentioned bandsaw geometry, we show how to resolve self-intersections and deform shapes so that they better tile space. \cref{sec:conclusions} discusses directions for future work. 

\section{Related Work}

\paragraph{Rasterization and Voxelization}

Rasterization of solid primitives such as meshes is one of the fundamental blocks of image generation, and techniques go back to the early days of computer graphics~\cite{halftonedrawings1967,shadingsolids1968,hiddensurface1969}. Generalization from two-dimensional pixels to three-dimensional voxels was made early on~\cite{volumeintegral1982}, enabling a structured geometry representation that allows for easier computation of e.g. translucent geometry, Constructive Solid Geometry (CSG) or inter-object intersections, which is usually easier to calculate over volumes than boundaries. 
Due to the regular structure and independence of the voxels as in the 2D case, algorithms have been developed that take advantage of the highly parallelization possibilities of the GPU~\cite{GPUVoxelization2008,solidvoxelGPU2010,crassin2012voxelization}.
The simplest form of rasterization/voxelization is point sampling the occupancy at the pixel/voxel centers. This leads to aliasing artifacts if frequencies higher than half the sampling
rate are present. Convolving the signal with a filter reduces aliasing, enables the use of the divergence theorem to iterate over the boundary instead of the volume~\cite{Wavelet2011} and allows differentiation~\cite{Sun2024}.

For our primal voxelization, we took inspiration from \citet{Rasterization2013}, as we use a 3D generalization of the scanline approach that handles overlapping shapes inherently and is optimized for meshes.

\paragraph{Volumetric Representations in Gradient-based Optimizations}

Volumetric shape representations, such as density~\cite{OccupancyNetworks2019, Chen2019} or signed distance functions~\cite{DeepSDF2019,SDFDiff2020,NeuS2021,learningsdf2021} are ubiquitous in modern computer vision and machine learning pipelines. Conversion between representations is common, but the prevailing direction is from volumetric to explicit~\cite{Liao2018, remelli2020meshsdf, binninger2025tetweave, shen2021dmtet, shen2023flexicubes}. This is partially motivated by the fact that some operations (e.g. rendering) or surface characteristics (e.g. developability) are more naturally expressed on the explicit representation.  Accordingly, our approach follows the \emph{inverse} direction by converting from explicit to volumetric representations in a fully differentiable manner. This is particularly useful for objectives that are more naturally expressed on the volume, such as resolving collisions during the optimization of articulated meshes~\cite{hassan2019resolving, jiang2020coherent, khirodkar2022occluded, Xu2025handshadow} or for using volume-based neural networks on such representations~\cite{maturana2015voxnet}.

\paragraph{Winding Numbers.}

The winding number~\cite{mobius1865bestimmung} has found many applications in geometry processing, including mesh booleans and CSG operations~\cite{mesharangement2016}, robust tetrahedral meshing~\cite{Jacobson2013,TetWild2018}, surface reconstruction from point clouds~\cite{fastwindingnumbers2018,WindPoly2024, Chen2024}, and mesh repair for non-watertight inputs~\cite{MeshRepair2013}.
 
The method of \citet{Jacobson2013}, provides a robust generalization of winding number that gracefully handles non-watertight meshes and triangle soups. \citet{fastwindingnumbers2018} accelerated this computation to $O(n \log n)$ using a hierarchical \citet{Barnes:1986nb} style algorithm, making winding number queries more practical for large meshes and multiple queries at once. %

Most closely related to our work, \citet{Sun2024} introduced Gaussian-smoothed winding numbers with analytic derivatives for surface and sampling point optimization. They solve a boundary integral for each voxel separately, made practical only in 2D by approximating the surface using a \citet{Barnes:1986nb} style simplification of the curve.
\citet{Sun2026} further used Gaussian-smoothed winding numbers in 3D, to guide the optimization of offset surfaces. In comparison, our key insight is that by using a box filter instead of a Gaussian, the regular grid structure of the sampling points can be exploited, leading to a highly efficient, differentiable voxelization algorithm without requiring acceleration data structures.

\section{Differentiable Voxelization}
\label{sec:differentiable_voxelization}

\emph{Voxelization} of a closed surface $\surface$ in $\R^n$ can generally be understood as point-sampling its winding number field $\windingsub(\mv{x}) \in \R$ on a regular grid
\begin{equation}
    \windingsub(\mv{x}_{\mv{i}}), %
\end{equation}
where $\mv{x}_{\mv{i}} \in \R^n$ is the center of the grid cell, or \emph{voxel}, with multi-index $\mv{i} = (i_1, \dots, i_n)_{1 \leq i_k \leq d_k}$. 
This perspective aligns with common intuition: if the surface is intersection-free, the winding number reduces to the indicator function
\begin{equation}
\label{eq:indicator}
    \indicator(\mv{x}) \coloneq \left\{ \begin{array}{cl}
        1, & \mv{x} \in \shape \\
        0, & \mv{x} \not\in \shape
    \end{array} \right.  
\end{equation}
of the region $\shape \subset \R^n$ bounded by the surface $\surface$. 
In this case, the value of each voxel indicates if its center $\mv{x}_{\mv{i}}$ is inside or outside of the represented shape.
In the following, we will omit the subscripts for brevity, such that $\winding(\mv{x})$ means $\windingsub(\mv{x})$.

Our goal is to differentiate the voxel representation with respect to parameters $\params \in \R^m$ that govern the surface $\surface$ (e.g., vertex positions of a mesh).
While the winding number at each voxel naturally depends on the surface, and therefore the parameters $\params$, the particular difficulty is that the winding number field is piecewise constant: the derivative $\dfdx{\winding(\mv{x})}{\params}$\footnote{We use $\dfdx{f}{\sceneparam}$ as shorthand notation for $\frac{\partial f}{\partial \params}$.} is zero almost everywhere on $\R^n$, so the derivative at the voxel centers $\dfdx{\winding(\mv{x}_{\mv{i}})}{\params}$ is either zero or undefined.
Instead, we consider the \emph{pre-filtered} winding number
\begin{equation}
\label{eq:winding_prefiltered}
    \windingsmooth(\point) = \int_{\R^n} \kernel(\point - \point') \winding(\point') \; \text{d}A(\point'), 
\end{equation}
where $\kernel$ is a filter kernel. For our particular choice of kernels, this corresponds to a smoothing of the winding number field, which leads to voxel values $\windingsmooth(\mv{x}_\mv{i})$ that continuously vary with the surface.
In \cref{sec:general_derivatives}, we will first derive an expression for the derivative $\dfdx{\windingsmooth(\point)}{\params}$ that is valid for general surfaces in arbitrary dimensions.
Without loss of generality, in most of our derivations, we will consider just a single surface parameter $\sceneparam \in \R$ for brevity.

The ability to differentiate the voxel representation of a surface enables gradient-based solutions of optimization problems
\begin{equation}
\label{eq:optimization_problem}
    \boldsymbol{\theta}^\star = \argmin_{\boldsymbol{\theta} \in \R^m} \loss(\mv{W}),
\end{equation}
where $\loss$ is some energy, or objective function, taking the voxel grid $\mv{W} \in \R^{d_1 \times \dots \times d_n}$, with $\mv{W}_\mv{i} = \windingsmooth(\mv{x}_\mv{i})$, as input.
\cref{sec:general_integral_estimators} investigates the steps required for such optimizations, in particular how to compute the smoothed winding numbers $\windingsmooth(\point_\mv{i})$ and their derivatives $\dfdx{\windingsmooth(\point_\mv{i})}{\params}$ in the general case.

We make the key observation that restricting the setting to one of the most common cases -- piecewise linear surfaces with box filters -- allows highly efficient voxelization and differentiation.
The respective algorithms are introduced in \cref{sec:closed_form_voxelization}.

\subsection{Derivatives for General Surfaces in $\R^n$}
\label{sec:general_derivatives}

For the 2D case ($n = 2$), \citet{Sun2024} differentiate \cref{eq:winding_prefiltered}, assuming that $\surface$ is a piecewise linear curve and $\kernel$ a Gaussian kernel, by locally decomposing the winding number into Heaviside step functions. 
We find that Reynold's transport theorem~\cite{Reynolds1903} enables a simple derivation for general surfaces in any dimension. This takes inspiration from differentiable rendering~\cite{Zhang2019}, where similar integrals have been differentiated to optimize scene geometry using image-based objectives\footnote{The given setting has two key differences to differentiable rendering: (1) the interior term of the derivative vanishes, leaving only the boundary term; and (2) the discontinuity points are not object silhouettes (i.e., subsets of the surface), but the \emph{entire} surface, which is trivial to sample.
}.

Consider the derivative of \cref{eq:winding_prefiltered} for a surface parameter $\sceneparam \in \R$:
\begin{equation}
    \dfdx{\windingsmooth(\point)}{\sceneparam} = \dfdx{}{\sceneparam} \int_{\R^n} \kernel(\point - \point') \winding(\point') \, \text{d}A(\point').
\end{equation}
It is incorrect to simply ``differentiate under the integral'' because the integrand has discontinuities in $\R^n$ that move with $\sceneparam$: the winding number $\winding$ is piecewise constant, and the constant regions are bounded by subsets of the surface $\surface$. In our case, the winding number field is also the \emph{only} source of such discontinuities, even if the kernel is discontinuous, because $\sceneparam$ only affects the surface and neither the kernel nor shape or position of the voxel grid.

The general idea is to partition the domain into those regions $\region_i \subset \R^n$ with constant winding number $\winding_i$
\begin{equation}
    = \sum_{i} \dfdx{}{\sceneparam} \int_{\region_i(\theta)} \kernel(\point - \point') \winding_i \, \text{d}A(\point').
\end{equation}
Since the region boundaries are subsets of the surface, they depend on $\sceneparam$. Applying Reynold's transport theorem splits the differential integral into an \emph{interior} and a \emph{boundary} term:
\definecolor{ColorInterior}{HTML}{A7C7E7}
\definecolor{ColorBoundary}{HTML}{F4A6A6}
\begin{multline}
    = \sum_{i} \mymathbox{ColorInterior}{Interior}{
        \int_{\region_i \setminus \partial \region_i} \dfdx{}{\sceneparam} \big[ \kernel(\point - \point') \winding_i \big] \, \text{d}A(\point')
    } \quad + \\
    \mymathbox{ColorBoundary}{Boundary}{
        \int_{\partial \region_i} \dfdx{\point_b}{\sceneparam}^\top \normal_i(\point_b) \; \kernel(\point - \point_b) \winding_i \; \text{d} \ell(\point_b)
    }.
\end{multline}
The boundary integral is over region boundaries $\partial \region_i$, where $\normal_i$ is the outward-pointing normal and $\dfdx{\point_b}{\sceneparam}^\top \normal_i$ measures the normal velocity of the boundary point $\point_b$ -- its variation with $\sceneparam$ in normal direction. The interior integrates the derivative of the integrand, which vanishes because the integrand is independent of $\sceneparam$. 
Therefore, the derivative simplifies to the boundary term:
\begin{equation}
    = \sum_{i} \int_{\partial \region_i} \dfdx{\point_b}{\sceneparam}^\top \normal_i(\point_b) \; \kernel(\point - \point_b) \winding_i \; \text{d}\ell(\point_b). \label{eq:derivative_boundary_split}
\end{equation}
Unfortunately, the region boundaries $\partial \region_i$ are unions of not necessarily connected subsets of $\surface$, which are difficult to construct explicitly if the surface has self-intersections. 
However, as we will show, this sum of integrals can be simplified to a single integral over the surface $\surface$ that is independent of winding numbers.

\paragraph{Special case: intersection-free surfaces}
\label{sec:method_nonintersect_surfaces}
Before we continue with the general case of possibly self-intersecting surfaces, consider the simplified case without intersections for an intuition. We only distinguish two regions in this case: (1) the inner region $\region_1 \coloneq \shape$ representing a solid shape bounded by its surface $\partial \region_1 \coloneq \surface$ where the normal $\normal_1 \coloneq \normal$ aligns with the normal of the surface $\normal$; and (2) the outer region $\region_2 \coloneq \R^n \setminus \shape $ with the same boundary $\partial \region_2 \coloneq \surface$ but \emph{inverted} normal $\normal_2 \coloneq -\normal$. Then \cref{eq:derivative_boundary_split} simplifies to
\begin{equation}
\label{eq:derivative_boundary_no_intersect}
    \dfdx{\windingsmooth}{\sceneparam}(\point) \stackrel{\footnotesize
        \substack{\makebox[0pt]{$\surface$ \text{without self-intersections}}\\[0.5ex]\displaystyle\downarrow\\ ~}
    }{=} 
    \int_{\surface} \dfdx{\point_b}{\sceneparam}^\top \normal \, \kernel(\point - \point_b) \; (\underbrace{\winding_1}_{\coloneq 1} - \underbrace{\winding_2}_{\coloneq 0}) \; \text{d}\ell(\point_b),
\end{equation}
since the winding number reduces to the indicator function for intersection-free surfaces (see \cref{eq:indicator}).

\paragraph{General case}
\label{sec:method_general_surfaces}

We now consider the general case of possibly self-intersecting surfaces. First note that the surface $\surface$ separates regions with piecewise constant winding number by definition\footnote{If the self-intersection has non-zero measure on the surface, %
the intersection creates a zero-volume region $\region_i$ with proper winding number.}, so let $\partial \region_{i, j}$ denote the shared boundary between $\region_i$ and $\region_j$. The union of shared boundaries is therefore simply the boundary itself
\begin{equation}
    \partial \region_i = \bigcup_{j \neq i} \partial  \region_{i, j}.
\end{equation}
Splitting the integration domain in \cref{eq:derivative_boundary_split} into shared boundaries gives
\begin{equation}
    \dfdx{\windingsmooth}{\sceneparam}(\point) = \sum_i \sum_{j \neq i} \int_{\partial \region_{i, j}} \dfdx{\point_b}{\sceneparam}^\top \normal_i(\point_b) \; \kernel(\point - \point_b) \winding_i \; \text{d} \ell(\point_b).
\end{equation}
Note that the shared boundaries are a double cover of the surface $\surface$: Each part is considered twice, once as $\partial \region_{i, j}$ with normal $\normal_i$ and once as $\partial \region_{j, i}$ with normal $\normal_j = -\normal_i$. Without loss of generality, assume that $\normal_i$ aligns with the surface normal $\normal$ on $\region_{i, j}$, then we can write the integral as
\begin{equation}
    = \sum_{\partial \region_{i, j}} \int_{\partial \region_{i, j}} \dfdx{\point_b}{\sceneparam}^\top \normal(\point_b) \; \kernel(\point - \point_b) (\winding_i - \winding_j) \; \text{d}\ell(\point_b).
\end{equation}
Since $\winding_i = \winding_j + 1$ by definition of the winding number, we have
\begin{equation}
\label{eq:derivative_boundary_final}
    \dfdx{\windingsmooth}{\sceneparam}(\point) = \int_{\surface} \dfdx{\point_b}{\sceneparam}^\top \normal(\point_b) \; \kernel(\point - \point_b) \; \text{d}\ell(\point_b),
\end{equation}
which is a surface integral independent of winding numbers.
This result is the same as that for non-intersecting surfaces (c.f. \cref{eq:derivative_boundary_no_intersect}).

\subsection{Optimizing with Integral Estimators}
\label{sec:general_integral_estimators}

\begin{figure}
    \centering
    \includegraphics[width=\linewidth, trim=0 20 0 0]{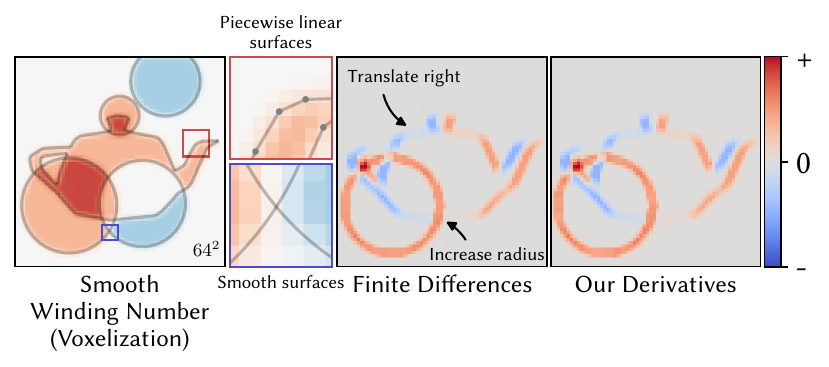}
    \caption{We can differentiate voxelizations, i.e., point-sam\-pled winding numbers, for arbitrary surfaces. This example shows piecewise linear surfaces and circles in 2D. Our derivatives match the finite difference reference. This example uses Monte Carlo integration and a box filter of diameter 3.}
    \label{fig:dvx_2d}
\end{figure}

Computing the gradient $\dfdx{E}{\sceneparam}$ to solve an optimization problem involving the voxel grid, as in \cref{eq:optimization_problem}, boils down to three steps:
(1) estimate the smoothed winding number field $\windingsmooth$ at the voxel centers $\mv{x}_\mv{i}$ with \cref{eq:winding_prefiltered}, (2) evaluate the energy $E$ to compute the \emph{adjoint} winding number $\dfdxfrac{E}{\windingsmooth(\mv{x}_\mv{i})}$ for each voxel, and (3) estimate the gradient %
\begin{equation}
    \dfdx{\loss}{\sceneparam} = \sum_\mv{i} \frac{\partial \loss}{\partial \windingsmooth(\mv{x}_\mv{i})} \dfdx{\windingsmooth}{\sceneparam}(\mv{x}_{\mv{i}})    
\end{equation}
as product of adjoints and winding number derivatives with \cref{eq:derivative_boundary_final}.

Steps (1) and (3) are unique to differentiable voxelization, and both require estimating integrals as shown in \cref{sec:general_derivatives}. 
Since we have so far not made any assumption about the surface $\surface$, the kernel $\kernel$, or the dimension, we will briefly discuss possible integration strategies for the general case (see \cref{fig:dvx_2d} for results). 
For the most common case in practice -- piecewise linear surfaces in 2D and 3D -- we introduce a highly efficient closed-form algorithm in \cref{sec:closed_form_voxelization}.

\paragraph{Smoothed winding number field}
\label{sec:general_estimator_for_w}

The smoothed winding number at voxel centers $\windingsmooth(\point_\mv{i})$ can be evaluated by estimating the integral in Eq.~\eqref{eq:winding_prefiltered}. A straightforward approach would be quadrature on $\R^n$ -- which, however, is prone to aliasing -- or Monte Carlo integration. As a simple baseline, we have implemented a Monte Carlo estimator of Eq.~\eqref{eq:winding_prefiltered} that uses stratified sampling with sub-voxel strata. The appeal of these direct approaches is that they can be trivially parallelized and use established rendering techniques because (1) samples in a voxel affect only their local neighborhood in the grid if the kernel $k$ is (approximately) compact, and (2) the winding number can be computed by counting signed ray intersections with the surface, using a \emph{single} ray in any direction.

The winding number computation can easily become a bottleneck in this approach. \citet{Sun2024} prove\footnote{Their setting is the 2D case, but the proof generalizes to arbitrary dimensions.} that one can eliminate the winding numbers from the integral by transforming \cref{eq:winding_prefiltered} into a surface integral using the divergence theorem:
\begin{equation}
\label{eq:winding_prefiltered_div}
    \windingsmooth(\point) = \int_{\surface} \mv{F}_\point^\top(\point_b) \normal(\point_b) \, \text{d} \ell(\point_b), 
\end{equation}
where $\mv{F}_\point$ is a vector field that satisfies $\nabla \cdot \mv{F}_\point(\point') = k(\point - \point')$. 
It seems appealing to estimate this integral by simply sampling the boundary and assigning the respective sample contributions to all voxels where $\mv{F}_{\point_\mv{i}}(\point_b) \neq 0$. 
However, this leads to visible ``bar''-shaped correlation artifacts in the voxel grid, which is in line with what Sun et al. observe in preliminary quadrature experiments for the 3D case \cite[Fig.14]{Sun2024}.
In 2D, their alternative is an artifact-free closed-form solution of \cref{eq:winding_prefiltered_div} for Gaussian kernels. 
We derive a similar closed-form solution of \cref{eq:winding_prefiltered_div} for piecewise linear surfaces in 2D and 3D (see \cref{sec:closed_form_voxelization}).
For the general case, we believe that more effective Monte Carlo estimators of \cref{eq:winding_prefiltered_div} can be derived, but this is left for future work.

\paragraph{Differential integral}

The gradient $\dfdx{\loss}{\sceneparam}$ is obtained by evaluating the product $\dfdxfrac{\loss}{\windingsmooth(\mv{x}_\mv{i})} \dfdx{\windingsmooth}{\sceneparam}(\mv{x}_{\mv{i}})$ for each voxel center $\mv{x}_\mv{i}$. 
This amounts to estimating the integral in \cref{eq:derivative_boundary_final} with the adjoint $\dfdxfrac{\loss}{\windingsmooth(\mv{x}_\mv{i})}$ 
moved inside of the integral. 
Similar to the winding number integral (\cref{sec:general_estimator_for_w}), we have implemented a simple Monte Carlo baseline for the derivative of general kernels and shapes. The derivative is localized to the surface $\surface$ for (approximately) compact kernels $k$, so we ``splat'' contributions of surface samples into the voxel grid instead of solving the integral for each voxel separately. 

\subsection{Efficient Closed-Form Integration}
\label{sec:closed_form_voxelization}

While the evaluation in the general setting may require appropriately adjusted quadrature schemes, we notice that certain restrictions on the kernel and the surface representation simplify matters and lead to a very efficient algorithm -- in particular for the practically relevant case of box filters and simplicial meshes. In this section, we will assume the 3D case. Given the explanation in 3D, the two-dimensional case or higher dimensions should be clear.

\paragraph{Winding numbers for piecewise surfaces}

We begin by recalling and formalizing important established properties of the winding number~
\cite{Grunbaum:1990:wn}: The winding number $w(\mv{x})$ at position $\mv{x}$ can be computed by summing the \emph{`signed intersections'} between the surface $\surface$ and a single ray from $\point$ into an arbitrary direction $\mv{d}_\mv{x}$: starting from zero, the winding number increases by one if $\mv{d}_\mv{x}\tp\mv{n} > 0$ and decreases by one if $\mv{d}_\mv{x}\tp\mv{n} < 0$, where $\mv{n}$ is the surface normal at the intersection.
If we consider a division of the surface into patches $p_i$, these intersections can be computed patch-wise
meaning the winding number at position $\mv{x}$ is the sum over all patch intersections:
\begin{equation}
\label{eq:patch_wise_winding}
    w(\mv{x}) = \mathrm{sgnInt}(\mv{x},\mv{d}_\mv{x},\surface) = \sum\nolimits_i \mathrm{sgnInt}(\mv{x},\mv{d}_\mv{x},p_i) =\sum\nolimits_i w_i(\mv{x},\mv{d}_\mv{x}),
    \vspace{0.5em}
\end{equation}
where $\mathrm{sgnInt}(\mv{x},\mv{d}_\mv{x},\surface)$ is the sum of `signed intersections' of the ray $\mv{x}+t\mv{d}_\mv{x}, \,t\in\mathbb{R}$ and the surface $\surface$, which is divided into patches $p_i$, where $p_i\subset\surface$ is an arbitrary disjunct patch division ($p_i\cap p_j=\emptyset$ for $i\neq j$) of the surface $\surface = \bigcup_i p_i$. This results in a `signed intersections' (or winding number contribution) field $w_i(\mv{x},\mv{d}_\mv{x})$ per patch. The idea is now to define a direction $\mv{d}_\mv{x}$ per point $\mv{x}$ in space and a patch division that allows us to compute $w_i$ efficiently.

\begin{figure}
    \centering
    \includegraphics[width=\linewidth]{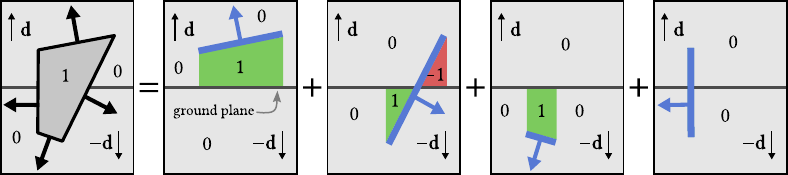}
    \caption{The winding number field $w$ of a surface depicted in the left-most box is the sum of the four patch-wise contributions $w_i$. Each $w_i$ defines for every point in space how often the patch $p_i$ (blue) is intersected in direction $\pm\mv{d}$. As the ray intersects the patch at most once, $w_i$ is piecewise constant and either $-1,1$ or $0$ depending on the normal relative to the ground plane.}
    \label{fig:segment_contribution}
\end{figure}

We make two choices: (1) we set $\mv{d}_\mv{x}$ to be constant; and (2) assume that each surface patch has at most one intersection with any ray in the direction $\mv{d}_x$. This is true, importantly, for triangles\footnote{Apart from degenerate cases, which can be ignored because their contribution will later become zero.}. Given the observation that now $w_i$ changes along a ray in direction $\mv{d}_\mv{x}$ only at an intersection with the surface patch $p_i$ from $-1$ or $1$ to $0$ and is $0$ everywhere else, we see that $w_i$ is piecewise constant and either $-1, 1$ or $0$. A complete example is depicted in \cref{fig:segment_contribution}.
For efficiency reasons, which we will describe later, we actually define two different directions $\mv{d}$ and $-\mv{d}$, pointing in opposite directions, divided by a \emph{ground plane} such that the direction in each half-space is orthogonal to the ground plane and points away from it. %

\paragraph{Box filtering on voxel grids}
\label{par:box_filter_voxel_grids}

So far, we considered the winding number and not the \emph{filtered} winding number. %
If we restrict the kernel to a box filter aligned with the voxels, the sample values for the voxels taken at their centers become:
\begin{equation}
\label{eq:winding_smooth_box}
    \windingsmooth(\point_\mv{i}) = \int_{\R^n} \text{box}(\point_\mv{i} - \point') \winding(\point') \, \text{d}A(\point') = \frac{1}{|\Box_{\point_\mv{i}}|}
    \int_{\Box_{\point_\mv{i}}} \winding(\point') \, \text{d}A(\point'),
\end{equation}
where $|\Box_{\point_\mv{i}}|$ is the volume of the voxel (i.e., of the box filter) and
\begin{equation}
    \Box_{\point_\mv{i}} = \left\{
\mv{y} \mid \point_{\mv{i}, j} - \frac{\mv{l}_j}{2} \le \mv{y}_j < \point_{\mv{i}, j} +\frac{\mv{l}_j}{2}, j \in [1,\ldots,n]
\right\},
\end{equation}
where $(\cdot)_j$ is the vector entry in dimension $j$ and $\mv{l}$ is the vector with the side lengths of the voxel. So $\windingsmooth$ evaluated at the voxel center is just the \emph{average} winding number inside the voxel. 

If a patch does not cross the ground plane, the corresponding $w_i$ takes only two values: $0$ and either $1$ or $-1$. In this case, we can interpret $w_i$ as the winding number field of the oriented \emph{shadow surface} of the patch.
The shadow surface bounds the \define{shadow solid}, which is the region between the patch and its `shadow' -- the (orthogonal) orthographic projection onto the ground plane (c.f. colored
area in \cref{fig:segment_contribution}).
The orientation of the shadow surface is consistent with the patch normal and determines the sign of the shadow solid volume.
Substituting \cref{eq:patch_wise_winding} into \cref{eq:winding_smooth_box}
shows that we can compute the average winding number inside a voxel by summing over all average winding numbers of all shadow surfaces. This is simply the signed volume of the intersection of the shadow solid with each voxel, divided by the volume of the voxel.

\begin{figure*}[htb]
    \centering
    \includegraphics[width=\linewidth]{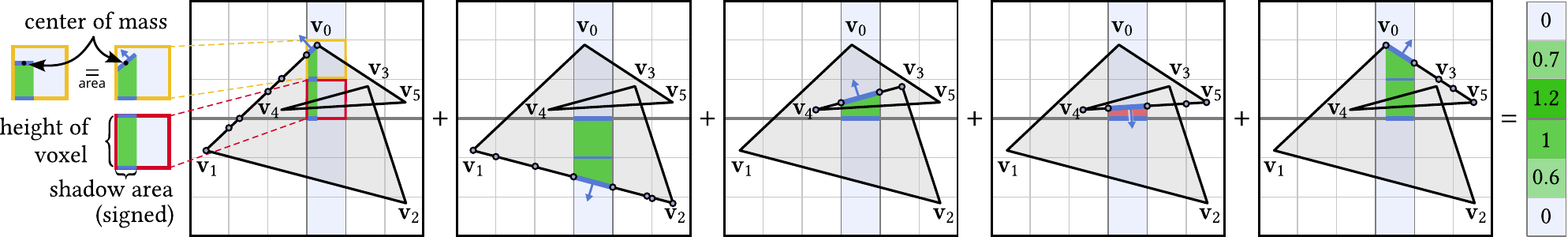}
    \caption{Given a polygon, the integrated winding numbers in each voxel are updated whenever a simplex intersects a voxel above the voxel. The volume of the intersection of the signed shadow solid volume (green for positive and red for negative contribution) and a voxel (small boxes) can be computed as described in \cref{par:box_filter_voxel_grids} and added to the value of each voxel, resulting in an update of all voxel values `below' the simplex segment. Therefore, the integrated winding number of all voxels in the highlighted grid column can be computed as the sum of `shadow' contributions of the five blue segments.}
    \label{fig:2d_explaination}
\end{figure*}

\paragraph{Computation for a triangle mesh} 
Assuming the surface is a triangle mesh, the general idea is to compute the intersection of each triangle with all voxels and re-triangulate the resulting polygons. This results in new sub-triangles that are fully contained within a voxel. 
If the ground plane is aligned with the voxel boundaries, no sub-triangle will cross it, making the \hyperlink{shadow solid}{shadow solid} perspective applicable.
As the shadow solid of each sub-triangle is a truncated prism that covers only voxels within a column of the grid, the signed volume of the voxel intersections can be easily computed:
\begin{enumerate}
    \item Choose a ground plane such that $\mv{d}$ is aligned with one of the grid axes and represent all points in this coordinate system. W.l.o.g., set $(x,y,0)$ as the ground plane and $\mv{d}=(0,0,1)$.
    
    \item For each triangle, compute the intersection with the voxel grid. 
    This gives convex polygons with degrees between 3 and 9 that are each fully contained in a single voxel $(i,j,k)$. Each polygon is further triangulated into sub-triangles by fixing a vertex and walking over the edges not incident on that vertex. This results in up to 7 sub-triangles per polygon.

    \item The non-zero contribution of each sub-triangle needs to be distributed to the voxel $(i,j,k)$ and all the voxels below, down to the ground plane at $(i,j,0)$. 
    
    \item[(3.1)] The contribution to voxel $(i,j,k)$ is the signed volume of the shadow solid, minus the contribution to the voxels below. The signed shadow solid volume is the signed area of the shadow of the patch on the ground plane -- the signed area of the triangle ignoring the third coordinate -- times the average height of the triangle over the ground plane -- the third coordinate of the centroid of the sub-triangle.
    
    \item[(3.2)] The contribution to each voxel below $(i,j,k)$ is the signed shadow area times the side length of each voxel in direction $\mv{d}$. This can be done efficiently for all voxels in a process very similar to the scan-conversion of primitives.
\end{enumerate}

\cref{fig:2d_explaination} visualizes this process in 2D.
This algorithm is a special case of the scanline algorithm of \citet{Rasterization2013} extended to 3D, who show that this method can also be extended to other filters. A pseudo-code with a ground plane at the bottom of the voxel grid is provided in the supplementary material.

\paragraph{Acceleration.}
In practice, we use a ground plane at the center of the voxel grid, which halves the average distance of voxels to the ground plane and reduces the number of write operations for meshes where most triangles are near the ground plane. Alternatively, it is possible to first gather the downward contributions of each voxel and then distribute them downward in a second pass.

\paragraph{Derivatives}
\label{sec:closed_form_derivatives}

Given the grid-sized box filter, by \cref{eq:derivative_boundary_final}, the derivative of the winding number at the voxel centers is
\begin{align}
    \dfdx{\windingsmooth}{\sceneparam}(\point_\mv{i}) &= \int_{\surface} \dfdx{\mv{x}_b}{\sceneparam}^\top \normal(\mv{x}_b) \; \text{box}(\mv{x} - \mv{x}_b) \; \text{d} \ell(\mv{x}_b)\\
     &= \frac{1}{|\Box_{\point_\mv{i}}|}\int_{\surface \cap \Box_{\point_\mv{i}}} \dfdx{\mv{x}_b}{\sceneparam}^\top \normal(\mv{x}_b) \; \text{d}\ell(\mv{x}_b).
\end{align}
For a piecewise linear mesh where all triangles $t_i$ are entirely within the voxel ($t_i = t_i \cap \Box_{\mv{x}}$), this reduces to
\begin{equation}\label{eq:diff_box_contained}
     \dfdx{\windingsmooth}{\sceneparam}(\point_\mv{i}) = \frac{1}{|\Box_{\point_\mv{i}}|}\sum_{t_i} \sum_{\mv{v}_{j}\in t_i}\frac{1}{3}\dfdx{\mv{v}_{j}}{\sceneparam}^\top\, \normal_{t_i} \; \text{Area}(t_i),
\end{equation}
where $\mv{v}_{j}$ are the three vertices of the triangle $t_i$. This is because the normal vector $\normal_{t_i}$ is constant over the triangle and the derivative of the surface over a triangle with respect to the vertices ($\dfdx{\mv{x}_b}{\mv{v}}$) is a linear function, going from 1 at the respective vertex to 0 at the opposite edge, making it integrable by taking the center of mass of the triangle times its area. If we now assume arbitrary triangles, possibly spanning multiple voxels, we can divide each triangle into sub-triangles $s_j \in (t_i \cap \Box_x)$, as described above. According to \cref{eq:diff_box_contained}, we can compute the derivative explicitly by
\begin{equation}
     \dfdx{\windingsmooth}{\sceneparam}(\point_\mv{i}) = \frac{1}{|\Box_{\point_\mv{i}}|}\sum_{(\vphantom{s_j\surface}\surface \ni t_i) \cap \Box_{\point_\mv{i}}} \sum_{\vphantom{s_j\surface}s_j \in (t_i \cap \Box_{\point_\mv{i}})}\sum_{\vphantom{s_j\surface}\mv{w}_{k} \in s_j}\frac{1}{3}\dfdx{\mv{w}_{k}}{\sceneparam}^\top \normal_{t_i} \; \text{Area}(s_j).
\end{equation}
The three vertices $\mv{w}_k$ of the sub-triangle $s_j$ are expressed as linear combinations of the original triangle $t_i$ vertices ($\mv{v}_j$ in \cref{eq:diff_box_contained}), ensuring the dependence on $\theta$.
Therefore, the derivatives can be computed using the same structure as the voxelization algorithm, without the downward contribution, if we represent the triangle-voxel intersections in barycentric coordinates. 
This can be seen by comparing the voxelization pseudocode Alg. 1 with the differentiation in Alg. 2 and Alg. 3 in the supplementary material.

\paragraph{Box filters of different sizes}

The box filter can be larger than a grid cell. 
The Monte Carlo implementation \cref{sec:general_integral_estimators} handles this out-of-the-box (see \cref{fig:dvx_2d}). 
For the closed-form integration, one can apply another discrete filter to the already box-filtered winding number, as computed above.
If the second filter is again a box filter, the composition of voxelization and discrete convolution is equivalent to using a single box filter with integer-scaled support.

\section{Evaluation and Applications}
\label{sec:application}

We implemented the algorithms using \textsc{C++} with Python bindings for PyTorch~\cite{Paszke2019}. %
We will first start by investigating the performance of our approach and comparing it with another differentiable voxelization technique. Afterwards, we demonstrate the versatility of our differentiable voxelization with applications for self-intersection resolving (\cref{sec:resolving_intersections}), optimization for bandsaw cutting (\cref{sec:bandsaw}), and space tiling using shapes in $\R^3$ (\cref{sec:space_tiling}).

\label{sec:implementation}
\label{sec:evaluation}

\subsection{Evaluation}

\paragraph{Runtime} 

\begin{figure}
    \centering
    \includegraphics[width=\linewidth]{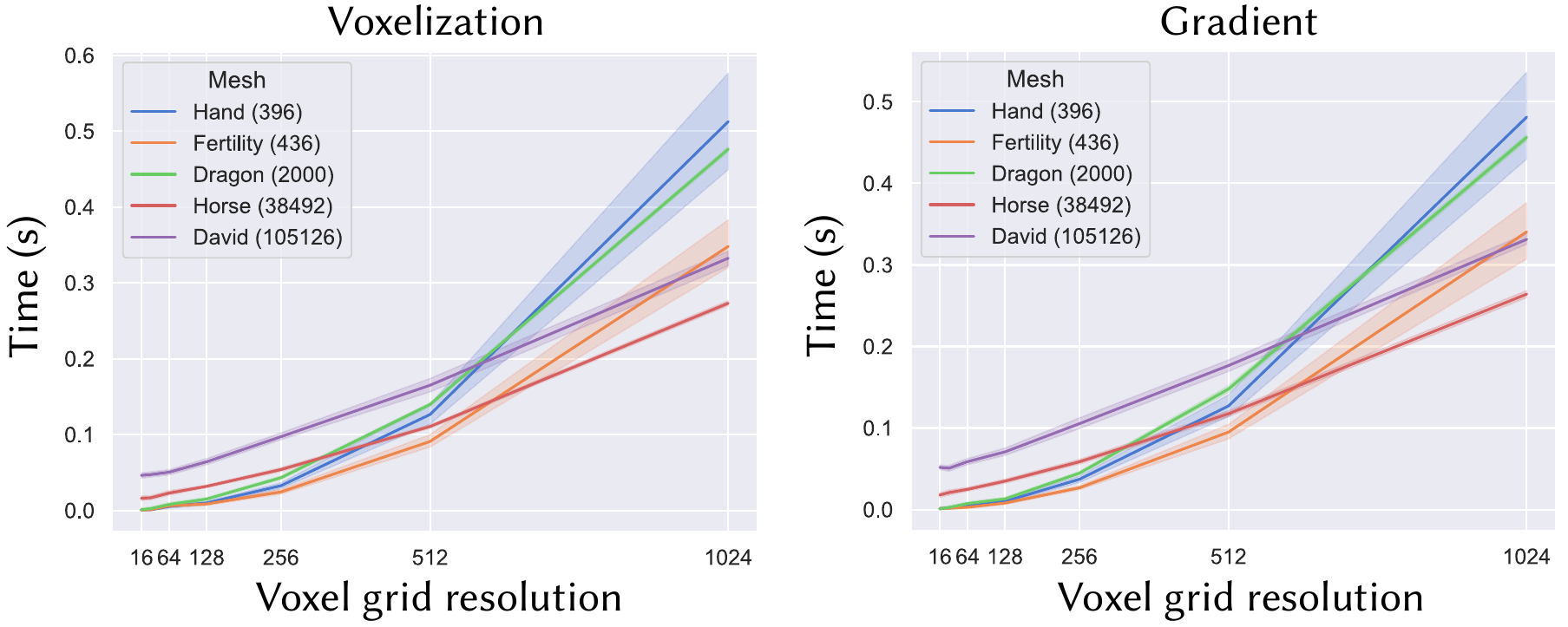}
    \caption{For high grid resolutions (up to $1024$ voxels in each dimension), the runtime stays below a second even for meshes over $100,000$ faces for voxelization as well as the gradient computation. All timings are averaged over $30$ repetitions and were computed on an Intel i9-12900K CPU. The number of faces is shown in parentheses.}
    \label{fig:timings}
\end{figure}

Computation time is a limiting factor for many volumetric representations, as the number of voxels grows cubically with the number of voxels in each dimension. We measure the computation time of our implementation in 32-bit floating-point precision on a 12th Gen Intel Core i9-12900K, for both the voxelization and the backward-mode gradient computation. We tested multiple meshes with different numbers of faces on grids between $16^3$ up to $1024^3$ voxels, by averaging the execution time over 30 runs per mesh and grid size (Fig~\ref{fig:timings}). Even though the implementation is not fully optimized (for example, it is not parallelized), the computation times stay below a second even for meshes with more than $100,000$ faces on $1024^3$ voxels. This makes it practical for optimizations with many gradient descent iterations. 
Interestingly, the overhead of the David mesh for low-resolution grids disappears for larger grid sizes. The reason for this is likely because the number of intersections needed to be computed for low-resolution meshes increases cubically, whereas for the David mesh it remains constant, since the triangles are typically contained within a voxel. It is very likely that, for even finer voxel grids, the David mesh will perform worse than simple meshes again.

Preliminary experiments with a naive GPU implementation indicate significant potential for performance improvements by exploiting parallelism %
(speedup of $3.6\times$
for voxelization and
$14.5\times$ for gradients, for meshes with more than $100,000$ faces on $1024^3$ grids). 
We leave an optimized GPU implementation for future work.

\paragraph{Comparison with DOPH} 

\begin{figure}
    \centering
    \includegraphics[width=0.95\linewidth]{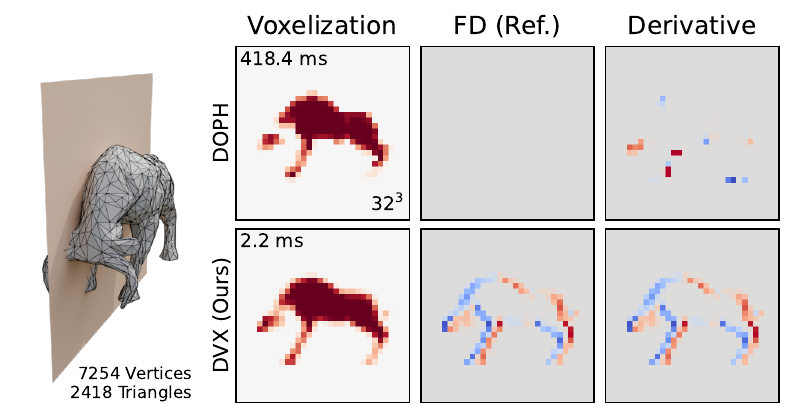}
    \caption{\citet{Luo2024differentiablevox} propose differentiable voxelization by automatically differentiating through a (non-smoothed) winding number (\textsc{DOPH}). Since the winding number is piecewise constant, the derivatives of their voxelization are zero if computed numerically with finite differences (FD), and their automatic derivatives are only spuriously non-zero very close to the surface. In contrast, our technique (\textsc{DVX}) is based on the smoothed winding number, which leads to expected numerical derivatives that match our results exactly.}
    \label{fig:dvx_vs_doph}
\end{figure}

Differentiable voxelization of surfaces has been attempted in previous work by \citet{Luo2024differentiablevox}: They point-sample the \emph{non-smoothed} winding number $\winding(\mv{x}_\mv{i})$ and use trilinear interpolation to get a smooth result. The derivative is then obtained with automatic differentiation. This approach corresponds to ``Discretize-then-Differentiate''~\cite{Bangaru2021}, which is known to produce incorrect derivatives: Since the winding number is piecewise constant, the derivative of their voxelization is zero almost everywhere. Our approach (Sec.~\ref{sec:differentiable_voxelization}) is based on point-sampling the smoothed winding number field $\windingsmooth(\mv{x}_\mv{i})$, which leads to the expected derivatives (see Fig.~\ref{fig:dvx_vs_doph}).

\subsection{Self-Intersection Resolving}
\label{sec:resolving_intersections}

\begin{figure}
    \centering
    \includegraphics[width=0.8\linewidth]{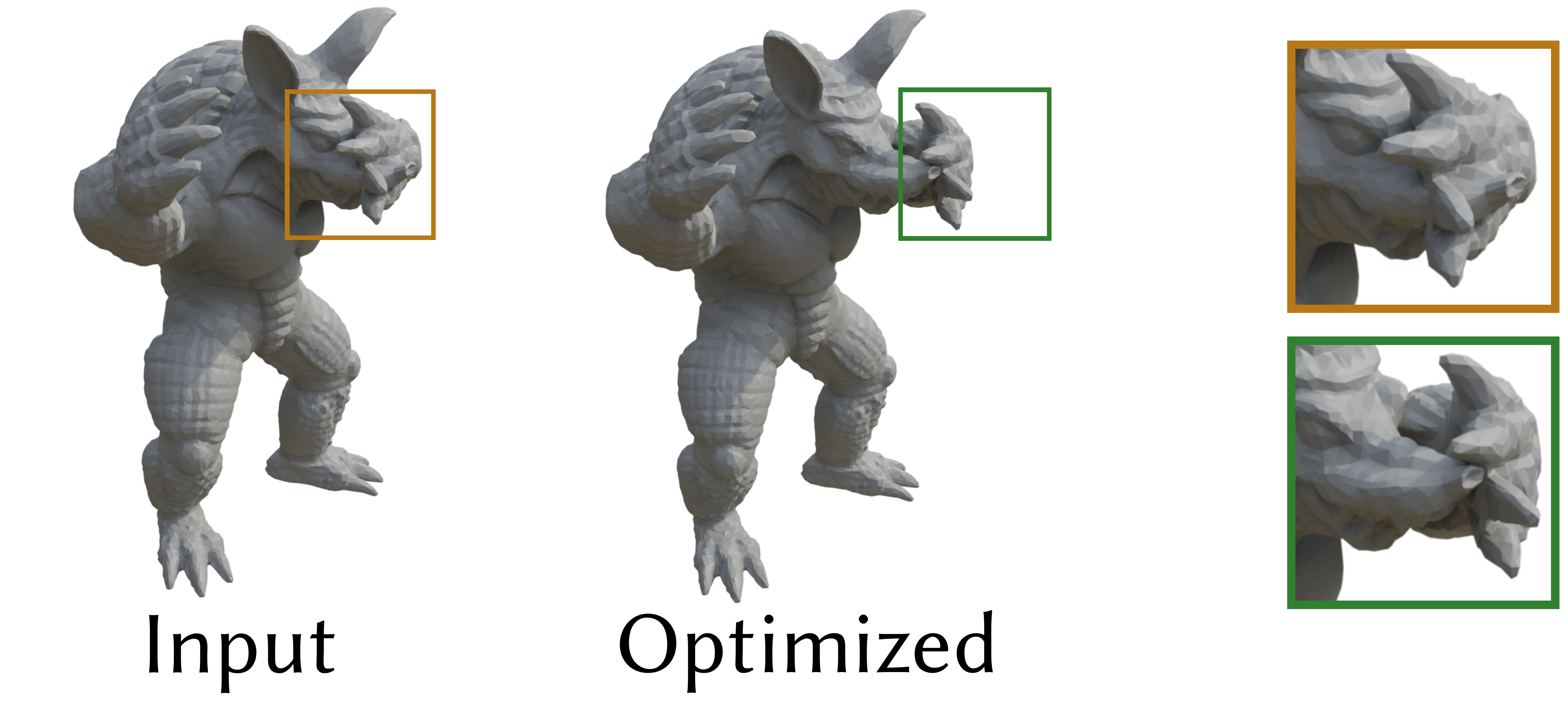}
    \caption{Differentiable voxelization allows resolving self-intersections by an ARAP deformation of the input shape.}
    \label{fig:resolved_intersections}
\end{figure}

Given a winding number, it is easy to detect self-intersections where multiple parts of the same surface overlap: these are all positions where the winding number is larger than one. Self-intersections may appear due to insufficient sampling of parametric models or as a result of many geometric optimization methods that do not penalize self-intersections such as ARAP surface deformation~\cite{Sorkine2007}. While \citet{Sun2024} showed that their method is suitable to resolve self-intersections for planar curves, our method can be used to resolve self-intersections for surfaces in $\R^3$.

Starting with a voxelization $\mv W \in \R^{d \times d \times d}$ of resolution $d$ of a triangle mesh ($\mv V$, $\mv F$), we use an energy functional that penalizes voxels with a value higher than one: $E_{\text{sect}} = \sum_{\mv W_{ijk}>1} \mv W_{ijk}^2$.
Due to numerical errors, it is advisable to filter values $\mv W_{ijk}>1 + \epsilon$ instead (for small $\epsilon = 0.001$) as some values of $\mv W$ can be larger than one.

Optimizing $E_{\text{sect}}$ using gradient descent %
resolves self-intersections in the input mesh. To prioritize rigid deformations of the input, we regularize the self-intersection energy using the as-rigid-as-possible (ARAP) energy $ E = E_{\text{sect}} + \alpha E_{\text{arap}}$~\cite{Sorkine2007} as it is often used as regularization for surface deformation~\cite{Cubic2019,ARAPReg2021,PDynamics2023,Style2025}:
\begin{equation}
    E_{\text{arap}}(\mv{\bar V}) = \min_{\mv R_i \in SO(3)} \sum_{i \in \mv V}^n \sum_{(j,k) \in \mathcal{N}(i)} w_{jk} ||\mv{e}_{jk} - \mv{R}_i \mv{\bar e}_{jk}||^2
    \label{eq:energy_arap}
\end{equation}
where $\mathcal{N}(i)$ are all spokes and rims half edges $\mv e_{jk} = \mv v_k - \mv v_j$ for vertex $i$ in the meshes~\cite{Chao2010}. Note that this energy is fully differentiable, and optimal rotations $\mv R_i$ between vertex stars can be found by singular value decompositions (for further details, we refer to Sorkine and Alexa~\shortcite{Sorkine2007}).

We minimize the regularized self-intersection energy with $\alpha = 0.005$ by gradient descent using the Adam method~\cite{Kingma2017} and a learning rate of $10^{-2}$. The optimization is iterated until all voxels with $\mv W_{ijk}>1$ are resolved (Fig.~\ref{fig:resolved_intersections}).

This experiment demonstrates that differentiable physical simulations that require cumbersome collision detection~\cite{meshphysics2020,neuralphysics2023} could benefit from our fast differential collision detection and resolution.
Note that self-intersections are resolved only approximately. There may be intersections in half-empty voxels that are not penalized by $E_{\text{sect}}$, as the energy only integrates the winding number over the entire voxel. Due to volume inversions, the winding number may also be negative. This could hide self-intersections in the optimization objective.

\subsection{Shape Deformation for Bandsaw Cutting}~\label{sec:bandsaw}

\begin{figure}
    \centering
    \includegraphics[width=0.35\linewidth]{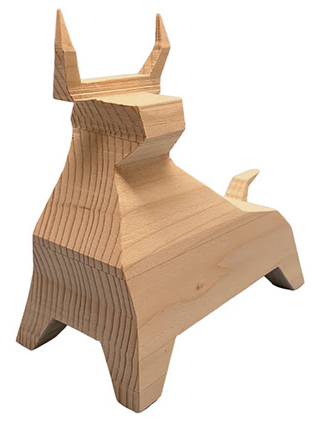}
    \includegraphics[width=0.49\linewidth]{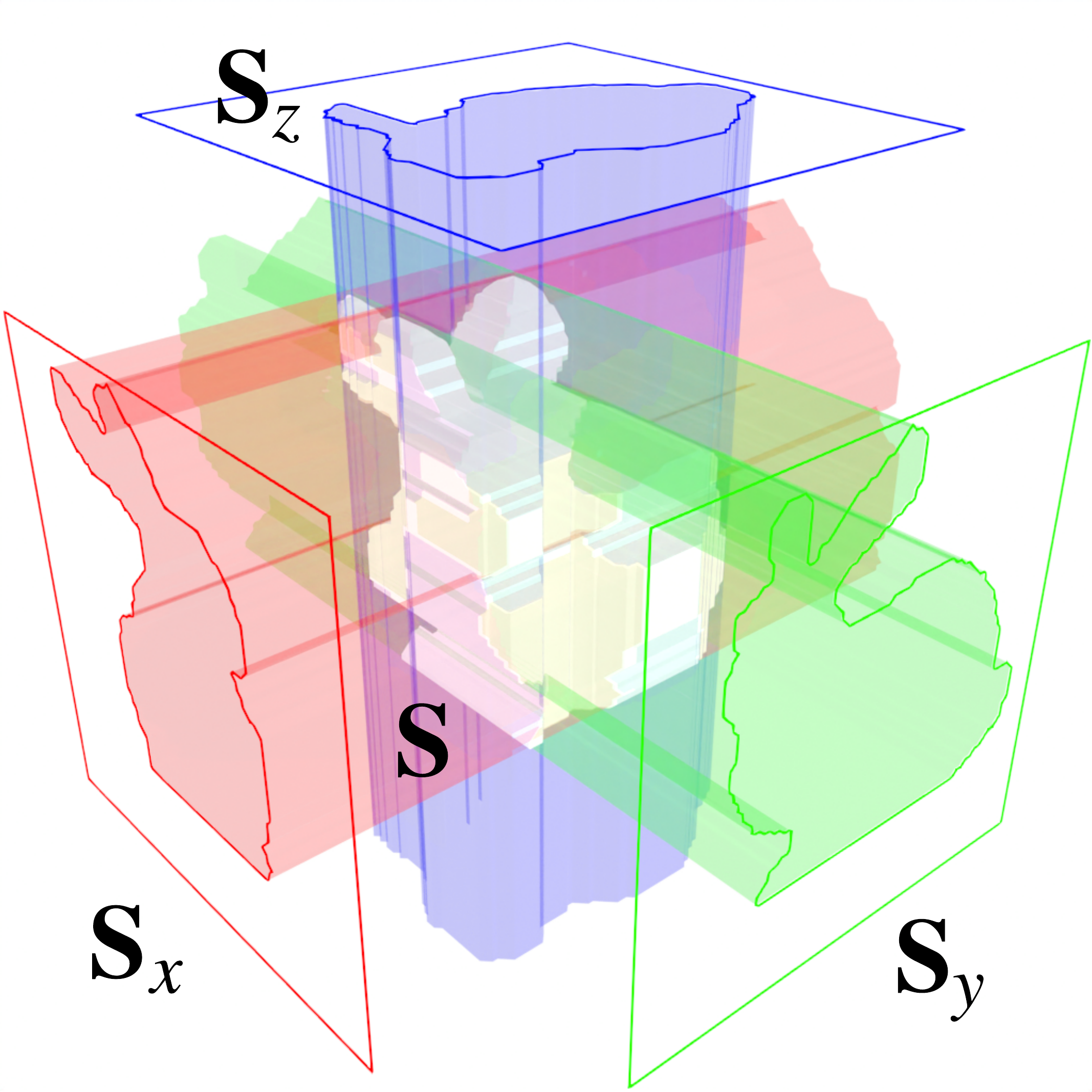}
    \caption{Left: Wooden figure that was created by cutting along handmade cut silhouettes~\cite{brikologik} (used with permission). Right: The cut out surface is defined as the intersection of extruded silhouettes.}
    \label{fig:bandsaw_art_and_silhouettes}
\end{figure}

\begin{figure*}
    \centering
    \includegraphics[width=0.9\linewidth]{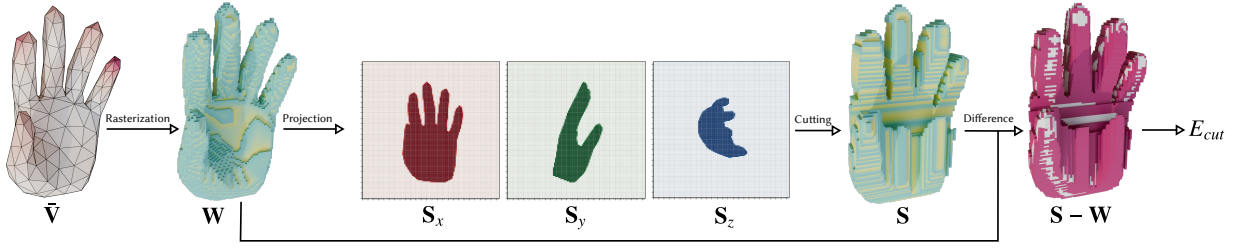}
    \caption{The energy $E_{cut}$ measures the volume difference between the deformed mesh $\mv{\bar V}$ and the deformed cut out shape $\mv S$. Larger vertex distances of the deformed mesh $\mv{\bar V}$ to $\mv V$ are shown in red (left). $\mv W$ is the discretized shape of $\mv{\bar V}$ where yellow voxels have high volume and green voxels have low volume. Voxels with volume lower than $10^{-3}$ are hidden. $\mv W$ is projected to the silhouette shapes $\mv S_x, \mv S_y$ and $\mv S_z$ that are used to generate the cut shape $\mv S$. For visualization of the volume difference ($\mv S - \mv W$), we highlight all voxels in red that are part of the shape $\mv S_{ijk} > 0.5$ but not in $\mv W_{ijk} > 0.5$.}
    \label{fig:cut_energy_overview}
\end{figure*}

\begin{figure*}
    \centering
    \includegraphics[width=\linewidth]{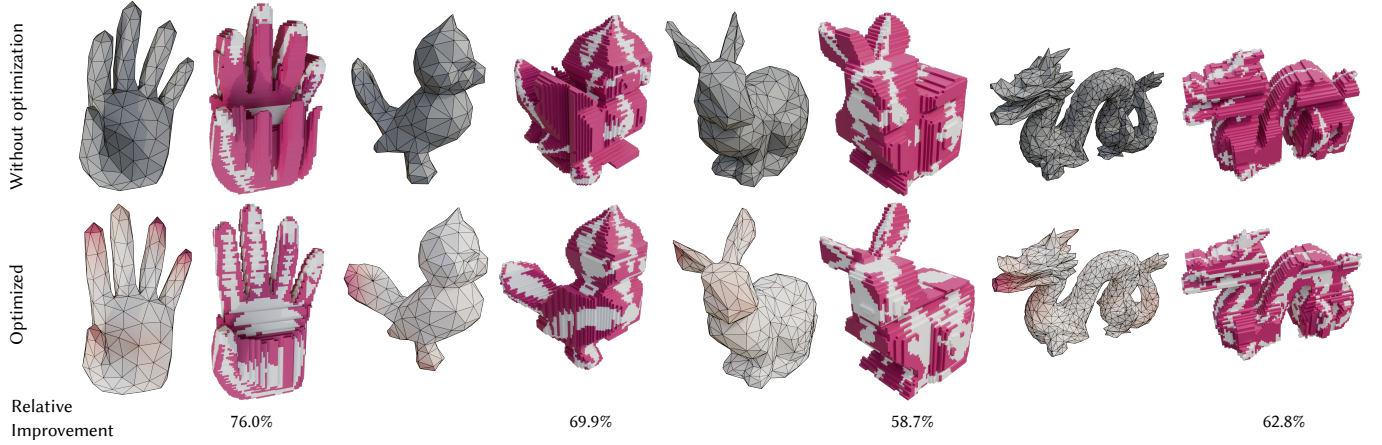}
    \caption{When optimized (bottom row), much more material that does not belong to the actual shape can be removed by cutting, when compared to the cut-out shape for meshes without optimization (top row). All voxels of $\mv S_{ijk} > 0.5$ that do not belong to $\mv W_{ijk} > 0.5$ are highlighted in red. The color of the optimized triangle meshes shows the difference to the initial meshes. Gray voxels are part of the voxelization of the triangle meshes and red voxels are additional voxels that are not removed by cutting. The relative improvements denote how much unnecessary material of the initial meshes can be removed with the optimization ($1-E_{cut}(\mv{\bar V})/ E_{cut}(\mv V)$). Minimal deformations, such as straightening the fingers of the hand or adjusting the tail and ears of Tweety and Bunny, significantly improve the shapes' cutability.}
    \label{fig:three_cut_directions}
\end{figure*}

Cutting objects from wooden blocks using orthogonal cutting directions restricts the space of shapes that can be created by sawing. For example, it is not possible to create certain types of concavities or surfaces with normals that are not orthogonal to at least one cut direction. Every \textit{cut out} surface is defined by a set of silhouettes, one for each cut direction. \citet{brikologik} offers such handmade cut silhouettes that result in beautiful wooden figures as depicted in Fig.~\ref{fig:bandsaw_art_and_silhouettes} (left).

A naive way to cut a given object is to directly use its silhouettes in $x$, $y$ and $z$ direction for sawing. This often leads to shapes that strongly differ from their initial shape due to material that is not removed but does not belong to the input shape (\cref{fig:three_cut_directions}, top row).
Our idea is to find small deformations of the input that make the shape more cutable but also preserve its overall shape.

We use a triangle mesh with vertices $\mv V$ and face indices $\mv F$  as input and compute the voxelized shape $\mv W \in \R^{d \times d \times d}$. We generate rasterized filled silhouettes $\mv S_x, \mv S_y$ and $\mv S_z \in \R^{d \times d}$ using the mellowmax operator~\cite{Asadi2017}, specifically $(\mv S_x)_{ij} =  1/w \ln( 1/d \sum_k \exp(w \mv W_{kij}))$.
For $w \rightarrow \infty$ the mellowmax converges to the maximum of the elements.
We use $w=1000$.
The gradient of the mellowmax is the softmax. If multiple voxels have the same value, then they contribute equally to the gradient.

The voxelized cut shape $\mv S \in \R^{d \times d \times d}$ is then given by product of the extruded silhouettes, i.e., $\mv S_{ijk} = (\mv S_x)_{jk} \ (\mv S_y)_{ik} \ (\mv S_z)_{ij}$ (Fig.~\ref{fig:bandsaw_art_and_silhouettes}, right).
The cut energy that measures the squared voxelized volume differences between the cut shape and the voxelized input is: $E_{cut} = \frac{1}{d^3} \sum_{ijk} (\mv S_{ijk} - \mv W_{ijk})^2$.

In Fig.~\ref{fig:cut_energy_overview} we visualize the steps to compute the energy function $E_{cut}$. Note that $\mv W$ contains winding numbers that are greater than one for self intersections. This intentionally penalizes self intersections that require additional volume. The final energy is the cut energy regularized by the as-rigid-as-possible energy $E = E_{cut} + \alpha E_{arap}$.
The parameter $\alpha$ weights between the additional volume and the deformation cost. We minimize the energy  with $\alpha = 0.05$ by gradient descent using the Adam method~\cite{Kingma2017} with a learning rate of $10^{-3}$.

We limit the resolution of the voxelization to $d=128$. When cutting a wooden block of $10\textrm{cm}^3$, this gives a resolution of $<0.8\textrm{mm}^3$, which is below the precision of many standard bandsaws.

\begin{figure}
    \centering
    \includegraphics[width=0.88\linewidth]{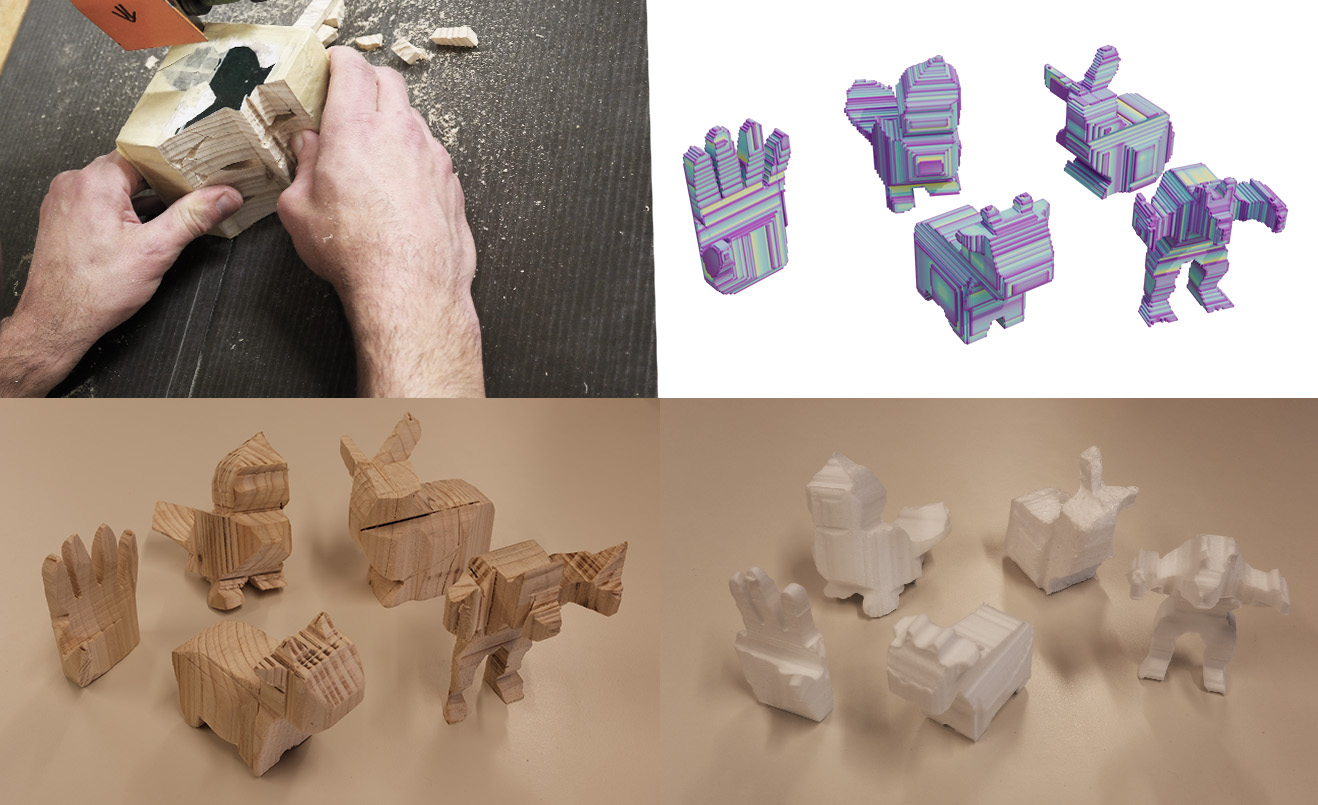}
    \caption{We test the `cutability' of optimized cut shapes by cutting various optimized shapes using a bandsaw (left top and bottom) and a polystyrene foam cutter (bottom right). The references for the cut shapes are at the top right.} %
    \label{fig:fabrication}
\end{figure}

In Fig.~\ref{fig:three_cut_directions} we show various meshes optimized for improved bandsaw cutting.
These deformations preserve the overall meaning of the shapes, but substantially improve the cut object when compared to the naive cutting strategy without deformation. We demonstrate the `cutability' of the generated surfaces by cutting various examples out of wood with a bandsaw and out of polystyrene foam with a foam cutter (Fig.~\ref{fig:fabrication}).

To quantify the improved volume difference, we applied the optimization to the TOSCA data set~\cite{Bronstein2008}.
For each optimized mesh, we compute the relative volume improvement between initial voxelized cut shape and cut shape for the voxelized optimized shape. We also compute the Hausdorff and mean surface distance between the initial and deformed triangle meshes using METRO~\cite{Cignoni1998} to quantify the deformation (Tab. 1 in the supplementary material).

\subsection{Space-filling Shapes in $\R^3$}\label{sec:space_tiling}

\begin{figure*}[htb]
    \includegraphics[width=\linewidth,trim={0 0 0 0},clip]{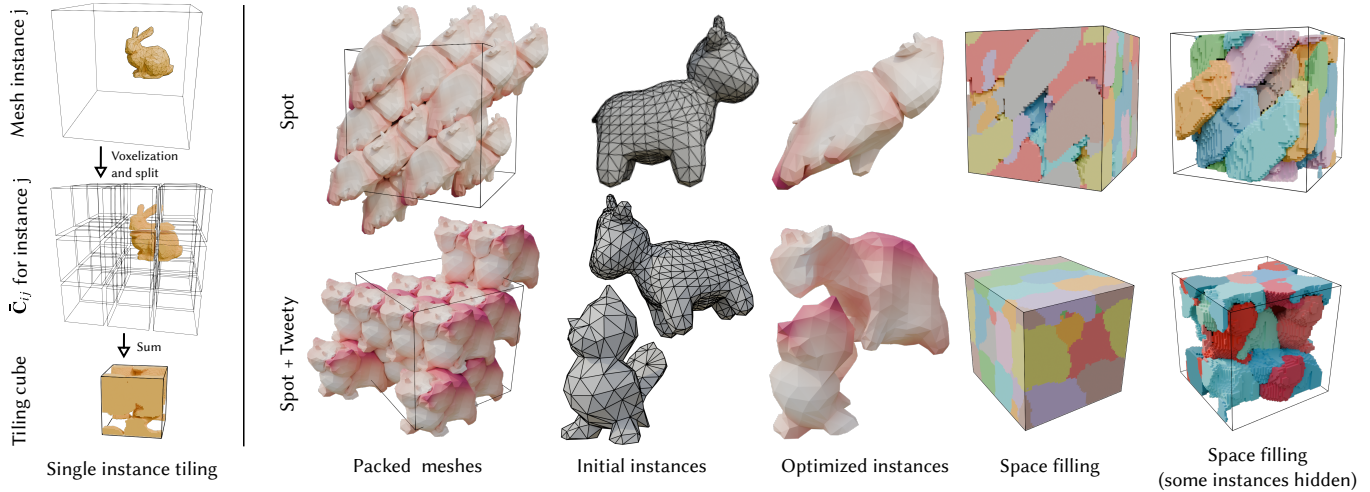}
    \caption{Differentiable voxelization can be used to optimize tilings in $\R^3$ to periodically fill the space. Left: Tiling procedure for a single shape instance. Right: Each row belongs to one optimized packing. The packed meshes (first column) can be repeated along all three axis to fill the space. The packed meshes are composed of one or more deformed meshes that are all similar (within a class of animals). When voxelized and repeated in all dimensions, the packing fills $\R^3$ (third column). The last column shows the interior of the space filling voxels by hiding some instances
    (for the space filling for multiple animal types with some hidden instances, all Tweety birds are shaded in reds and all Spots are shaded in blues).}
    \label{fig:escher}
\end{figure*}

Inspired by the tessellations of M.C. Escher (see~\citet{Kaplan2000} for comparable examples), we can use differentiable voxelization to generate space-filling shapes in $\R^3$. Given a set of shapes, the goal is to find a deformation such that the voxelized shapes fill the entire space without overlapping. Escher created his tilings by drawing textures for various known tilings such that the textures connect continuously with adjacent tiles. The textures contained figures of one or more animals that are repeated within the texture. Based on this idea, we optimize repeated shapes in a cube such that shapes at the boundary of the cube continuously connect to shapes on the opposite boundary. All shapes inside the cube should be identical, up to translation and rotation, and all repetitions (\textit{instances}) of a shape should be similar.

For practical applications, this can be seen as instance of a packing problem. The goal is to find a small deformation for a given shape that maximizes the number of shapes that fit into a domain.

Outputs of the optimization are the deformed vertex positions for each shape, as well as a translation and rotation (parametrized by yaw, pitch and roll) for each instance of a shape. For initialization, we randomly position each shape multiple times inside the cube such that each mean of vertices is inside $[-\frac{1}{3}, \frac{1}{3}]^3$. We refer to this cube as \textit{tiling cube}, which can later be repeated to fill $\R^3$. This initialization might cause some shapes to stick out of the tiling cube. The idea is that everything outside should enter the cube on the opposite site to create a space-filling that is a continuous transitions between opposite sides of the cube. To achieve this, we voxelize the entire $[-1,1]^3$ cube for each mesh $j$ using the differential voxelizer on a voxel grid $\mv C_j \in \R^{d \times d \times d}$. This  $[-1,1]^3$ cube is split in $3^3=27$ smaller cubes $\mv{\bar C}_{ji} \in \R^{d/3 \times d/3 \times d/3}$ of side length $\frac{1}{3}$ (\cref{fig:escher}, left). The sum of all splitted cubes $\sum_i \mv{\bar C_{ji}}$ gives a repeatable tiling for each shape. We optimized the differentiable energy that measures the coverage of the space  by summation over all shapes and splitted cubes $\sum_{ij} \mv{\bar C_{ji}}$:
\begin{equation*}
    E_{\text{fill}} = \sum_{klm} \bigg(\Big(\sum_{ij} \mv{\bar C}_{ji}\Big)_{klm} - 1\bigg)^2.
\end{equation*}
This energy becomes zero when each voxel in the tiling cube is fully covered by one shape without self intersections. %
For the final energy, we again use the ARAP energy $E_{\text{arap}}$ as regularization: $E = E_{\text{fill}} + \alpha \sum_j E_{\text{arap}}(\mv{\bar V}_j)$.

Note that in practice $E_{\text{arap}}$ only needs to be computed for each class of shapes once in each iteration. We minimize the regularized energy by gradient descent with $\alpha = 0.005$ using the Adam method~\cite{Kingma2017} and a learning rate of $10^{-3}$.

We use a resolution of $d=192$. When split into multiple cubes, it creates a tileable cube with a resolution of $d/3 = 64$. The stopping criterion is reached when the change in all vertex positions is less than $10^{-3}$. We optimize a tileable cube filling with 27 Spots (\cref{fig:escher}, right side and top row) and 27 hands (\cref{fig:teaser}). We also experimented with two meshes, Tweety and Spot, to optimize a tileable cube filled with $14$ Tweety birds and $14$ Spots (\cref{fig:escher} right side and bottom row). Although the meshes do not fill the entire volume and may not be intersection-free, the experiments show near tileable results.

\section{Discussion and Future Work}
\label{sec:conclusions}
\label{sec:discussion}
\label{sec:futurework}

There are operations for which the regular grid structure significantly simplifies the computation. Nevertheless, surface representations are often the more prevalent representation in comparison to volumetric grids, and many methods specifically rely on triangular meshes, including the classical ARAP energy. This work transfers the simplicity of many problems that can be solved on voxel grids to common explicit surface representations. Optimizing cut shapes (Sec.~\ref{sec:bandsaw}) or space-fillings in $\R^3$ are non-trivial and challenging problems and we are unaware of any other methods apart from voxelization that could handle these problems.

Differentiable voxelization opens many possibilities for applications that benefit from voxel grids, which we could not cover in the application section (Sect.~\ref{sec:application}). This includes (1) applications in CSG, as intersections, differences, and unions of shapes are particularly simple to compute on grids. 
Morphological operators (2) can also be easily formulated on grids. Differentiable voxelization provides a method to differentiate through these operations back to a triangular mesh. This might inspire further applications for thin shell modeling. Many applications in machine learning (3) benefit from grid structures. This opens the possibility of directly comparing shapes in $\R^3$ in a simple differentiable way, instead of relying on image-based losses for shapes in $\R^3$~\cite{Yuan2024}.

Interestingly, the inverse problem of generating a surface mesh from a voxel representation has been solved many times using differentiable algorithms~\cite{MeshSDF2020}. However, our direction has not received much attention. Closing this loop efficiently enables the generation of self-supervised optimization to improve algorithms anywhere within the loop.

Besides applications that could benefit from differentiable voxelization, we see several possibilities to decrease computation time by parallelization. Although this work focuses mainly on triangular meshes, this approach can be further extended and evaluated for more surface representations or different smoothing filters that are fast to integrate, such as polynomials~\cite{Rasterization2013}.

\begin{acks}

The authors would like to thank Maximilian Kohlbrenner for the helpful discussions on winding numbers and for sharing source code, and the anonymous reviewers for their valuable comments.
This work was funded by the European Research Council (ERC) under the European Union’s Horizon 2020 research and innovation program (Grant agreement No. 101055448, ERC Advanced Grand EMERGE).
\end{acks}

\bibliographystyle{ACM-Reference-Format}
\bibliography{bibliography}

@String{Computing = "Computing" }

@String{Computer = "{IEEE} Computer" }

@String{AMS = "American Mathematical Society" }

@String{Springer = "Springer-Verlag" }

@book{Reynolds1903,
    title = {The sub-mechanics of the universe},
    author = {Reynolds, Osborne and Brightmore, Arthur William and Moorby, William Henry},
    volume = {3},
    year = {1903},
    publisher = {University Press}
}

@article{Zhang2019,
    title = {A differential theory of radiative transfer},
    author = {Zhang, Cheng and Wu, Lifan and Zheng, Changxi and Gkioulekas, Ioannis and Ramamoorthi, Ravi and Zhao, Shuang},
    journal = TOG_SIGGRAPHAsia,
    volume = {38},
    number = {6},
    pages = {227},
    year = {2019},
    publisher = {ACM}
}

@article{Jacobson2013,
author = {Jacobson, Alec and Kavan, Ladislav and Sorkine-Hornung, Olga},
title = {Robust inside-outside segmentation using generalized winding numbers},
year = {2013},
issue_date = {July 2013},
publisher = {Association for Computing Machinery},
address = {New York, NY, USA},
volume = {32},
number = {4},
issn = {0730-0301},
Nourl = {https://doi.org/10.1145/2461912.2461916},
doi = {10.1145/2461912.2461916},
journal = {ACM Trans. Graph.},
month = jul,
articleno = {33},
numpages = {12}
}

@book{Bronstein2008,
  title={Numerical geometry of non-rigid shapes},
  author={Bronstein, Alexander M and Bronstein, Michael M and Kimmel, Ron},
  year={2008},
  publisher={Springer Science \& Business Media}
}

@article{Cignoni1998,
author = {Cignoni, P. and Rocchini, C. and Scopigno, R.},
title = {Metro: Measuring Error on Simplified Surfaces},
journal = {Computer Graphics Forum},
volume = {17},
number = {2},
pages = {167-174},
doi = {https://doi.org/10.1111/1467-8659.00236},
NOurl = {https://onlinelibrary.wiley.com/doi/abs/10.1111/1467-8659.00236},
NOeprint = {https://onlinelibrary.wiley.com/doi/pdf/10.1111/1467-8659.00236},
year = {1998}
}

@inproceedings{Sun2024,
author = {Sun, Haoran and Wang, Jingkai and Bao, Hujun and Huang, Jin},
title = {GauWN: Gaussian-smoothed Winding Number and its Derivatives},
year = {2024},
isbn = {9798400711312},
publisher = {Association for Computing Machinery},
address = {New York, NY, USA},
NOurl = {https://doi.org/10.1145/3680528.3687569},
doi = {10.1145/3680528.3687569},
booktitle = {SIGGRAPH Asia 2024 Conference Papers},
articleno = {89},
numpages = {9},
location = {Tokyo, Japan},
series = {SA '24}
}

@article{Luo2024differentiablevox,
  title={Differentiable Voxelization and Mesh Morphing},
  author={Luo, Yihao and Wang, Yikai and Xiang, Zhengrui and Xiu, Yuliang and Yang, Guang and Yap, ChoonHwai},
  journal={arXiv preprint arXiv:2407.11272},
  year={2024}
}

@inproceedings{Sorkine2007,
  title={As-rigid-as-possible surface modeling},
  author={Sorkine, Olga and Alexa, Marc},
  booktitle={Symposium on Geometry processing},
  volume={4},
  pages={109--116},
  year={2007}
}

@misc{Kingma2017,
      title={Adam: A Method for Stochastic Optimization}, 
      author={Diederik P. Kingma and Jimmy Ba},
      year={2017},
      eprint={1412.6980},
      archivePrefix={arXiv},
      primaryClass={cs.LG}
}

@article{Chao2010,
author = {Chao, Isaac and Pinkall, Ulrich and Sanan, Patrick and Schr\"{o}der, Peter},
title = {A simple geometric model for elastic deformations},
year = {2010},
issue_date = {July 2010},
publisher = {Association for Computing Machinery},
address = {New York, NY, USA},
volume = {29},
number = {4},
issn = {0730-0301},
NOurl = {https://doi.org/10.1145/1778765.1778775},
doi = {10.1145/1778765.1778775},
journal = {ACM Trans. Graph.},
month = jul,
articleno = {38},
numpages = {6}
}

@InProceedings{Asadi2017,
  title = 	 {An Alternative Softmax Operator for Reinforcement Learning},
  author =       {Kavosh Asadi and Michael L. Littman},
  booktitle = 	 {Proceedings of the 34th International Conference on Machine Learning},
  pages = 	 {243--252},
  year = 	 {2017},
  editor = 	 {Precup, Doina and Teh, Yee Whye},
  volume = 	 {70},
  series = 	 {Proceedings of Machine Learning Research},
  month = 	 {06--11 Aug},
  publisher =    {PMLR},
  NOurl = 	 {https://proceedings.mlr.press/v70/asadi17a.html},
}

@inproceedings{Kaplan2000,
author = {Kaplan, Craig S. and Salesin, David H.},
title = {Escherization},
year = {2000},
isbn = {1581132085},
publisher = {ACM Press/Addison-Wesley Publishing Co.},
address = {USA},
NOurl = {https://doi.org/10.1145/344779.345022},
doi = {10.1145/344779.345022},
booktitle = {Proceedings of the 27th Annual Conference on Computer Graphics and Interactive Techniques},
pages = {499–510},
numpages = {12},
series = {SIGGRAPH '00}
}

@incollection{Paszke2019,
title = {PyTorch: An Imperative Style, High-Performance Deep Learning Library},
author = {Paszke, Adam and Gross, Sam and Massa, Francisco and Lerer, Adam and Bradbury, James and Chanan, Gregory and Killeen, Trevor and Lin, Zeming and Gimelshein, Natalia and Antiga, Luca and Desmaison, Alban and Kopf, Andreas and Yang, Edward and DeVito, Zachary and Raison, Martin and Tejani, Alykhan and Chilamkurthy, Sasank and Steiner, Benoit and Fang, Lu and Bai, Junjie and Chintala, Soumith},
booktitle = {Advances in Neural Information Processing Systems 32},
pages = {8024--8035},
year = {2019},
publisher = {Curran Associates, Inc.},
NOurl = {http://papers.neurips.cc/paper/9015-pytorch-an-imperative-style-high-performance-deep-learning-library.pdf}
}

@inproceedings{ARAPReg2021,
author = {Huang, Qixing and Huang, Xiangru and Sun, Bo and Zhang, Zaiwei and Jiang, Junfeng},
year = {2021},
month = {10},
pages = {5795-5805},
title = {ARAPReg: An As-Rigid-As Possible Regularization Loss for Learning Deformable Shape Generators},
doi = {10.1109/ICCV48922.2021.00576}
}

@inbook{PDynamics2023,
author = {Bouaziz, Sofien and Martin, Sebastian and Liu, Tiantian and Kavan, Ladislav and Pauly, Mark},
title = {Projective Dynamics: Fusing Constraint Projections for Fast Simulation},
year = {2023},
isbn = {9798400708978},
publisher = {Association for Computing Machinery},
address = {New York, NY, USA},
edition = {1},
nourl = {https://doi.org/10.1145/3596711.3596794},
booktitle = {Seminal Graphics Papers: Pushing the Boundaries, Volume 2},
articleno = {82},
numpages = {11}
}

@article{Cubic2019,
author = {liu, Hsueh-Ti Derek and Jacobson, Alec},
title = {Cubic stylization},
year = {2019},
issue_date = {December 2019},
publisher = {Association for Computing Machinery},
address = {New York, NY, USA},
volume = {38},
number = {6},
issn = {0730-0301},
url = {https://doi.org/10.1145/3355089.3356495},
doi = {10.1145/3355089.3356495},
journal = {ACM Trans. Graph.},
month = nov,
articleno = {197},
numpages = {10}
}

@InProceedings{Style2025,
  author    = {Dinh, Nam Anh and Lang, Itai and Kim, Hyunwoo and Stein, Oded and Hanocka, Rana},
  title     = {Geometry in Style: 3D Stylization via Surface Normal Deformation},
  booktitle = {Proceedings of the IEEE/CVF Conference on Computer Vision and Pattern Recognition (CVPR)},
  month     = {June},
  year      = {2025},
  pages     = {28456-28467}
}

@book{botsch2010polygon,
  title={Polygon mesh processing},
  author={Botsch, Mario and Kobbelt, Leif and Pauly, Mark and Alliez, Pierre and L{\'e}vy, Bruno},
  year={2010},
  publisher={CRC press}
}

@book{farin2002curves,
  title={Curves and surfaces for CAGD: a practical guide},
  author={Farin, Gerald E},
  year={2002},
  publisher={Morgan Kaufmann}
}

@book{warren2001subdivision,
  title={Subdivision methods for geometric design: A constructive approach},
  author={Warren, Joe and Weimer, Henrik},
  year={2001},
  publisher={Elsevier}
}

@inproceedings{Yuan2024,
author = {Yuan, Haocheng and Bousseau, Adrien and Pan, Hao and Zhang, Quancheng and Mitra, Niloy J. and Li, Changjian},
title = {DiffCSG: Differentiable CSG via Rasterization},
year = {2024},
isbn = {9798400711312},
publisher = {Association for Computing Machinery},
address = {New York, NY, USA},
NOurl = {https://doi.org/10.1145/3680528.3687608},
doi = {10.1145/3680528.3687608},
booktitle = {SIGGRAPH Asia 2024 Conference Papers},
articleno = {9},
numpages = {10},
location = {Tokyo, Japan},
series = {SA '24}
}

@article{Deng_Shi_Li_Zhou_Zhang_Li_2021, title={Voxel R-CNN: Towards High Performance Voxel-based 3D Object Detection}, volume={35}, url={https://ojs.aaai.org/index.php/AAAI/article/view/16207}, DOI={10.1609/aaai.v35i2.16207}, abstractNote={Recent advances on 3D object detection heavily rely on how the 3D data are represented, i.e., voxel-based or point-based representation. Many existing high performance 3D detectors are point-based because this structure can better retain precise point positions. Nevertheless, point-level features lead to high computation overheads due to unordered storage. In contrast, the voxel-based structure is better suited for feature extraction but often yields lower accuracy because the input data are divided into grids. In this paper, we take a slightly different viewpoint --- we find that precise positioning of raw points is not essential for high performance 3D object detection and that the coarse voxel granularity can also offer sufficient detection accuracy. Bearing this view in mind, we devise a simple but effective voxel-based framework, named Voxel R-CNN. By taking full advantage of voxel features in a two-stage approach, our method achieves comparable detection accuracy with state-of-the-art point-based models, but at a fraction of the computation cost. Voxel R-CNN consists of a 3D backbone network, a 2D bird-eye-view (BEV) Region Proposal Network, and a detect head. A voxel RoI pooling is devised to extract RoI features directly from voxel features for further refinement. Extensive experiments are conducted on the widely used KITTI Dataset and the more recent Waymo Open Dataset. Our results show that compared to existing voxel-based methods, Voxel R-CNN delivers a higher detection accuracy while maintaining a real-time frame processing rate, i.e., at a speed of 25 FPS on an NVIDIA RTX 2080 Ti GPU. The code is available at https://github.com/djiajunustc/Voxel-R-CNN.}, number={2}, journal={Proceedings of the AAAI Conference on Artificial Intelligence}, author={Deng, Jiajun and Shi, Shaoshuai and Li, Peiwei and Zhou, Wengang and Zhang, Yanyong and Li, Houqiang}, year={2021}, month={May}, pages={1201-1209} }

@ARTICLE{1646334,
  author={Voelcker, H.B. and Requicha, A.A.G.},
  journal={Computer}, 
  title={Geometric Modeling of Mechanical Parts and Processes}, 
  year={1977},
  volume={10},
  number={12},
  pages={48-57},
  doi={10.1109/C-M.1977.217601}}

@inproceedings{10.2312:egtp.19871000,
booktitle = {EG 1987-Technical Papers},
editor = {},
title = {{A Fast Voxel Traversal Algorithm for Ray Tracing}},
author = {Amanatides, John and Woo, Andrew},
year = {1987},
publisher = {Eurographics Association},
ISSN = {1017-4656},
ISBN = {},
DOI = {10.2312/egtp.19871000}
}

@article{Rasterization2013,
author = {Manson, Josiah and Schaefer, Scott},
title = {Analytic Rasterization of Curves with Polynomial Filters},
journal = {Computer Graphics Forum},
volume = {32},
number = {2pt4},
pages = {499-507},
doi = {https://doi.org/10.1111/cgf.12070},
NOurl = {https://onlinelibrary.wiley.com/doi/abs/10.1111/cgf.12070},
NOeprint = {https://onlinelibrary.wiley.com/doi/pdf/10.1111/cgf.12070},
year = {2013}
}

@article{Wavelet2011,
author = {Manson, J. and Schaefer, S.},
title = {Wavelet Rasterization},
journal = {Computer Graphics Forum},
volume = {30},
number = {2},
pages = {395-404},
doi = {https://doi.org/10.1111/j.1467-8659.2011.01887.x},
NOurl = {https://onlinelibrary.wiley.com/doi/abs/10.1111/j.1467-8659.2011.01887.x},
eprint = {https://onlinelibrary.wiley.com/doi/pdf/10.1111/j.1467-8659.2011.01887.x},
year = {2011}
}

@inproceedings{GPUVoxelization2008,
author = {Eisemann, Elmar and D\'{e}coret, Xavier},
title = {Single-pass GPU solid voxelization for real-time applications},
year = {2008},
isbn = {9781568814230},
publisher = {Canadian Information Processing Society},
address = {CAN},
booktitle = {Proceedings of Graphics Interface 2008},
pages = {73–80},
numpages = {8},
location = {Windsor, Ontario, Canada},
series = {GI '08}
}

@article{solidvoxelGPU2010,
author = {Schwarz, Michael and Seidel, Hans-Peter},
title = {Fast parallel surface and solid voxelization on GPUs},
year = {2010},
issue_date = {December 2010},
publisher = {Association for Computing Machinery},
address = {New York, NY, USA},
volume = {29},
number = {6},
issn = {0730-0301},
Nourl = {https://doi.org/10.1145/1882261.1866201},
doi = {10.1145/1882261.1866201},
journal = {ACM Trans. Graph.},
month = dec,
articleno = {179},
numpages = {10}
}

@inproceedings{shadingsolids1968,
author = {Appel, Arthur},
title = {Some techniques for shading machine renderings of solids},
year = {1968},
isbn = {9781450378970},
publisher = {Association for Computing Machinery},
address = {New York, NY, USA},
Nourl = {https://doi.org/10.1145/1468075.1468082},
doi = {10.1145/1468075.1468082},
booktitle = {Proceedings of the April 30--May 2, 1968, Spring Joint Computer Conference},
pages = {37–45},
numpages = {9},
location = {Atlantic City, New Jersey},
series = {AFIPS '68 (Spring)}
}

@phdthesis{hiddensurface1969,
author = {Warnock, John Edward},
title = {A hidden surface algorithm for computer generated halftone pictures},
year = {1969},
publisher = {The University of Utah},
note = {AAI6919002}
}

@inproceedings{halftonedrawings1967,
author = {Wylie, Chris and Romney, Gordon and Evans, David and Erdahl, Alan},
title = {Half-tone perspective drawings by computer},
year = {1967},
isbn = {9781450378963},
publisher = {Association for Computing Machinery},
address = {New York, NY, USA},
Nourl = {https://doi.org/10.1145/1465611.1465619},
doi = {10.1145/1465611.1465619},
booktitle = {Proceedings of the November 14-16, 1967, Fall Joint Computer Conference},
pages = {49–58},
numpages = {10},
location = {Anaheim, California},
series = {AFIPS '67 (Fall)}
}

@article{volumeintegral1982,
author = {Lee, Yong Tsui and Requicha, Aristides A. G.},
title = {Algorithms for computing the volume and other integral properties of solids. II. A family of algorithms based on representation conversion and cellular approximation},
year = {1982},
issue_date = {Sept 1982},
publisher = {Association for Computing Machinery},
address = {New York, NY, USA},
volume = {25},
number = {9},
issn = {0001-0782},
Nourl = {https://doi.org/10.1145/358628.358648},
doi = {10.1145/358628.358648},
journal = {Commun. ACM},
month = sep,
pages = {642–650},
numpages = {9}
}

@ARTICLE{Botsch:deformation,
  author={Botsch, Mario and Sorkine, Olga},
  journal={IEEE Transactions on Visualization and Computer Graphics}, 
  title={On Linear Variational Surface Deformation Methods}, 
  year={2008},
  volume={14},
  number={1},
  pages={213-230},
  doi={10.1109/TVCG.2007.1054}}

@inproceedings{NEURIPS2020_b4b75896,
 author = {Liu, Lingjie and Gu, Jiatao and Zaw Lin, Kyaw and Chua, Tat-Seng and Theobalt, Christian},
 booktitle = {Advances in Neural Information Processing Systems},
 editor = {H. Larochelle and M. Ranzato and R. Hadsell and M.F. Balcan and H. Lin},
 pages = {15651--15663},
 publisher = {Curran Associates, Inc.},
 title = {Neural Sparse Voxel Fields},
 url = {https://proceedings.neurips.cc/paper_files/paper/2020/file/b4b758962f17808746e9bb832a6fa4b8-Paper.pdf},
 volume = {33},
 year = {2020}
}

@InProceedings{Sun_2022_CVPR,
    author    = {Sun, Cheng and Sun, Min and Chen, Hwann-Tzong},
    title     = {Direct Voxel Grid Optimization: Super-Fast Convergence for Radiance Fields Reconstruction},
    booktitle = {Proceedings of the IEEE/CVF Conference on Computer Vision and Pattern Recognition (CVPR)},
    month     = {June},
    year      = {2022},
    pages     = {5459-5469}
}

@article{MONTAGNAT20011023,
title = {A review of deformable surfaces: topology, geometry and deformation},
journal = {Image and Vision Computing},
volume = {19},
number = {14},
pages = {1023-1040},
year = {2001},
issn = {0262-8856},
doi = {https://doi.org/10.1016/S0262-8856(01)00064-6},
url = {https://www.sciencedirect.com/science/article/pii/S0262885601000646},
author = {J. Montagnat and H. Delingette and N. Ayache}
}

@article{fastwindingnumbers2018,
author = {Barill, Gavin and Dickson, Neil G. and Schmidt, Ryan and Levin, David I. W. and Jacobson, Alec},
title = {Fast winding numbers for soups and clouds},
year = {2018},
issue_date = {August 2018},
publisher = {Association for Computing Machinery},
address = {New York, NY, USA},
volume = {37},
number = {4},
issn = {0730-0301},
Nourl = {https://doi.org/10.1145/3197517.3201337},
doi = {10.1145/3197517.3201337},
journal = {ACM Trans. Graph.},
month = jul,
articleno = {43},
numpages = {12}
}

@article{mesharangement2016,
author = {Zhou, Qingnan and Grinspun, Eitan and Zorin, Denis and Jacobson, Alec},
title = {Mesh arrangements for solid geometry},
year = {2016},
issue_date = {July 2016},
publisher = {Association for Computing Machinery},
address = {New York, NY, USA},
volume = {35},
number = {4},
issn = {0730-0301},
url = {https://doi.org/10.1145/2897824.2925901},
doi = {10.1145/2897824.2925901},
journal = {ACM Trans. Graph.},
month = jul,
articleno = {39},
numpages = {15}
}

@article{Bangaru2021,
    author = {Bangaru, Sai Praveen and Michel, Jesse and Mu, Kevin and Bernstein, Gilbert and Li, Tzu-Mao and Ragan-Kelley, Jonathan},
    title = {Systematically differentiating parametric discontinuities},
    year = {2021},
    issue_date = {August 2021},
    publisher = {Association for Computing Machinery},
    address = {New York, NY, USA},
    volume = {40},
    number = {4},
    issn = {0730-0301},
    url = {https://doi.org/10.1145/3450626.3459775},
    doi = {10.1145/3450626.3459775},
    journal = {ACM Trans. Graph.},
    month = jul,
    articleno = {107},
    numpages = {18}
}

@inproceedings{MeshSDF2020,
author = {Remelli, Edoardo and Lukoianov, Artem and Richter, Stephan R. and Guillard, Beno\^{\i}t and Bagautdinov, Timur and Baque, Pierre and Fua, Pascal},
title = {MeshSDF: differentiable iso-surface extraction},
year = {2020},
isbn = {9781713829546},
publisher = {Curran Associates Inc.},
address = {Red Hook, NY, USA},
booktitle = {Proceedings of the 34th International Conference on Neural Information Processing Systems},
articleno = {1884},
numpages = {11},
location = {Vancouver, BC, Canada},
series = {NIPS '20}
}

@inproceedings{Chen2019,
  title={Learning implicit fields for generative shape modeling},
  author={Chen, Zhiqin and Zhang, Hao},
  booktitle={Proceedings of the IEEE/CVF conference on computer vision and pattern recognition},
  pages={5939--5948},
  year={2019}
}

@inproceedings{Liao2018,
  title={Deep marching cubes: Learning explicit surface representations},
  author={Liao, Yiyi and Donne, Simon and Geiger, Andreas},
  booktitle={Proceedings of the IEEE conference on computer vision and pattern recognition},
  pages={2916--2925},
  year={2018}
}

@article{remelli2020meshsdf,
  title={Meshsdf: Differentiable iso-surface extraction},
  author={Remelli, Edoardo and Lukoianov, Artem and Richter, Stephan and Guillard, Benoit and Bagautdinov, Timur and Baque, Pierre and Fua, Pascal},
  journal={Advances in Neural Information Processing Systems},
  volume={33},
  pages={22468--22478},
  year={2020}
}

@article{binninger2025tetweave,
  title={TetWeave: Isosurface Extraction using On-The-Fly Delaunay Tetrahedral Grids for Gradient-Based Mesh Optimization},
  author={Binninger, Alexandre and Wiersma, Ruben and Herholz, Philipp and Sorkine-Hornung, Olga},
  journal={ACM Transactions on Graphics (TOG)},
  volume={44},
  number={4},
  pages={1--19},
  year={2025},
  publisher={ACM New York, NY, USA}
}

@inproceedings{shen2021dmtet,
    title = {Deep Marching Tetrahedra: a Hybrid Representation for High-Resolution 3D Shape Synthesis},
    author = {Tianchang Shen and Jun Gao and Kangxue Yin and Ming-Yu Liu and Sanja Fidler},
    year = {2021},
    booktitle = {Advances in Neural Information Processing Systems (NeurIPS)}
}

@inproceedings{maturana2015voxnet,
  title={Voxnet: A 3d convolutional neural network for real-time object recognition},
  author={Maturana, Daniel and Scherer, Sebastian},
  booktitle={2015 IEEE/RSJ international conference on intelligent robots and systems (IROS)},
  pages={922--928},
  year={2015},
  organization={Ieee}
}

@article{shen2023flexicubes,
  title={Flexible isosurface extraction for gradient-based mesh optimization},
  author={Shen, Tianchang and Munkberg, Jacob and Hasselgren, Jon and Yin, Kangxue and Wang, Zian and Chen, Wenzheng and Gojcic, Zan and Fidler, Sanja and Sharp, Nicholas and Gao, Jun},
  journal={ACM Transactions on Graphics (TOG)},
  volume={42},
  number={4},
  pages={1--16},
  year={2023},
  publisher={ACM New York, NY, USA}
}

@inproceedings{hassan2019resolving,
  title={Resolving 3D human pose ambiguities with 3D scene constraints},
  author={Hassan, Mohamed and Choutas, Vasileios and Tzionas, Dimitrios and Black, Michael J},
  booktitle={Proceedings of the IEEE/CVF international conference on computer vision},
  pages={2282--2292},
  year={2019}
}

@inproceedings{jiang2020coherent,
  title={Coherent reconstruction of multiple humans from a single image},
  author={Jiang, Wen and Kolotouros, Nikos and Pavlakos, Georgios and Zhou, Xiaowei and Daniilidis, Kostas},
  booktitle={Proceedings of the IEEE/CVF conference on computer vision and pattern recognition},
  pages={5579--5588},
  year={2020}
}

@inproceedings{khirodkar2022occluded,
  title={Occluded human mesh recovery},
  author={Khirodkar, Rawal and Tripathi, Shashank and Kitani, Kris},
  booktitle={Proceedings of the IEEE/CVF conference on computer vision and pattern recognition},
  pages={1715--1725},
  year={2022}
}

@article{Xu2025handshadow,
author = {Xu, Hao and Wang, Yinqiao and Mitra, Niloy J. and Liu, Shuaicheng and Heng, Pheng-Ann and Fu, Chi-Wing},
title = {Hand-Shadow Poser},
year = {2025},
issue_date = {August 2025},
publisher = {Association for Computing Machinery},
address = {New York, NY, USA},
volume = {44},
number = {4},
issn = {0730-0301},
url = {https://doi.org/10.1145/3730836},
doi = {10.1145/3730836},
journal = {ACM Trans. Graph.},
month = jul,
articleno = {153},
numpages = {16}
}

@incollection{crassin2012voxelization,
	title         = {Octree-Based Sparse Voxelization Using The GPU Hardware Rasterizer},
	author        = {Crassin, Cyril and Green, Simon},
	editor        = {Patrick Cozzi and Christophe Riccio},
	booktitle     = {OpenGL Insights},
	publisher     = {CRC Press},
	month         = {July},
	year          = {2012},
	doi           = {10.1201/b12288},
}

@article{neuralphysics2023,
author = {Cleac'h, Simon and Yu, Hong-Xing and Guo, Michelle and Howell, Taylor and Gao, Ruohan and Wu, Jiajun and Manchester, Zac and Schwager, Mac},
year = {2023},
month = {05},
pages = {1-8},
title = {Differentiable Physics Simulation of Dynamics-Augmented Neural Objects},
volume = {PP},
journal = {IEEE Robotics and Automation Letters},
doi = {10.1109/LRA.2023.3257707}
}

@inproceedings{meshphysics2020,
author = {Qiao, Yi-Ling and Liang, Junbang and Koltun, Vladlen and Lin, Ming C.},
title = {Scalable differentiable physics for learning and control},
year = {2020},
publisher = {JMLR.org},
booktitle = {Proceedings of the 37th International Conference on Machine Learning},
articleno = {727},
numpages = {10},
series = {ICML'20}
}

@INPROCEEDINGS{DeepSDF2019,
  author={Park, Jeong Joon and Florence, Peter and Straub, Julian and Newcombe, Richard and Lovegrove, Steven},
  booktitle={2019 IEEE/CVF Conference on Computer Vision and Pattern Recognition (CVPR)}, 
  title={DeepSDF: Learning Continuous Signed Distance Functions for Shape Representation}, 
  year={2019},
  volume={},
  number={},
  pages={165-174},
  doi={10.1109/CVPR.2019.00025},
  ISSN={2575-7075},
  month={June},}

@INPROCEEDINGS{OccupancyNetworks2019,
  author={Mescheder, Lars and Oechsle, Michael and Niemeyer, Michael and Nowozin, Sebastian and Geiger, Andreas},
  booktitle={2019 IEEE/CVF Conference on Computer Vision and Pattern Recognition (CVPR)}, 
  title={Occupancy Networks: Learning 3D Reconstruction in Function Space}, 
  year={2019},
  volume={},
  number={},
  pages={4455-4465},
  doi={10.1109/CVPR.2019.00459},
  ISSN={2575-7075},
  month={June},}

@INPROCEEDINGS{SDFDiff2020,
  author={Jiang, Yue and Ji, Dantong and Han, Zhizhong and Zwicker, Matthias},
  booktitle={2020 IEEE/CVF Conference on Computer Vision and Pattern Recognition (CVPR)}, 
  title={SDFDiff: Differentiable Rendering of Signed Distance Fields for 3D Shape Optimization}, 
  year={2020},
  volume={},
  number={},
  pages={1248-1258},
  doi={10.1109/CVPR42600.2020.00133},
  ISSN={2575-7075},
  month={June},}

@inproceedings{NeuS2021,
author = {Wang, Peng and Liu, Lingjie and Liu, Yuan and Theobalt, Christian and Komura, Taku and Wang, Wenping},
title = {NeuS: learning neural implicit surfaces by volume rendering for multi-view reconstruction},
year = {2021},
isbn = {9781713845393},
publisher = {Curran Associates Inc.},
address = {Red Hook, NY, USA},
booktitle = {Proceedings of the 35th International Conference on Neural Information Processing Systems},
articleno = {2081},
numpages = {13},
series = {NIPS '21}
}

@inproceedings{learningsdf2021,
author = {Zhang, Jingyang and Yao, Yao and Quan, Long},
year = {2021},
month = {10},
pages = {6505-6514},
title = {Learning Signed Distance Field for Multi-view Surface Reconstruction},
doi = {10.1109/ICCV48922.2021.00646}
}

@article{mobius1865bestimmung,
  title={{\"U}ber die Bestimmung des Inhaltes eines Poly{\"e}ders},
  author={Mobius, A},
  journal={Mathematisch-Physische Klasse},
  volume={17},
  pages={31--68},
  year={1865}
}

@article{MeshRepair2013,
author = {Attene, Marco and Campen, Marcel and Kobbelt, Leif},
title = {Polygon mesh repairing: An application perspective},
year = {2013},
issue_date = {February 2013},
publisher = {Association for Computing Machinery},
address = {New York, NY, USA},
volume = {45},
number = {2},
issn = {0360-0300},
nourl = {https://doi.org/10.1145/2431211.2431214},
doi = {10.1145/2431211.2431214},
journal = {ACM Comput. Surv.},
month = mar,
articleno = {15},
numpages = {33}
}

@article{TetWild2018,
author = {Hu, Yixin and Zhou, Qingnan and Gao, Xifeng and Jacobson, Alec and Zorin, Denis and Panozzo, Daniele},
title = {Tetrahedral meshing in the wild},
year = {2018},
issue_date = {August 2018},
publisher = {Association for Computing Machinery},
address = {New York, NY, USA},
volume = {37},
number = {4},
issn = {0730-0301},
nourl = {https://doi.org/10.1145/3197517.3201353},
doi = {10.1145/3197517.3201353},
journal = {ACM Trans. Graph.},
month = jul,
articleno = {60},
numpages = {14}
}

@inproceedings{WindPoly2024,
author = {He, Xin and Lv, Chenlei and Huang, Pengdi and Huang, Hui},
title = {WindPoly: Polygonal Mesh Reconstruction via Winding Numbers},
year = {2024},
isbn = {978-3-031-72969-0},
publisher = {Springer-Verlag},
address = {Berlin, Heidelberg},
nourl = {https://doi.org/10.1007/978-3-031-72970-6_17},
doi = {10.1007/978-3-031-72970-6_17},
booktitle = {Computer Vision – ECCV 2024: 18th European Conference, Milan, Italy, September 29–October 4, 2024, Proceedings, Part XLVII},
pages = {294–311},
numpages = {18},
location = {Milan, Italy}
}

@online{brikologik,
  author = {Pierre-Luc Gauthier},
  title = {Brikologik},
  year = 2019,
  url = {https://brikologik.com/en/collections/all},
  urldate = {2025-01-22}
}

@article{Barnes:1986nb,
    author = "Barnes, Josh and Hut, Piet",
    title = "{A Hierarchical 0 Nlogn Force Calculation Algorithm}",
    reportNumber = "PRINT-86-0299 (IAS,PRINCETON)",
    doi = "10.1038/324446a0",
    journal = "Nature",
    volume = "324",
    pages = "446--449",
    year = "1986"
}

@ARTICLE{Grunbaum:1990:wn,
  title     = "Rotation and winding numbers for planar polygons and curves",
  author    = "Gr{\"u}nbaum, Branko and Shephard, G C",
  journal   = "Trans. Am. Math. Soc.",
  publisher = "American Mathematical Society (AMS)",
  volume    =  322,
  number    =  1,
  pages     = "169--187",
  year      =  1990,
  language  = "en"
}

@ARTICLE{Sun2026,
  author={Sun, Haoran and Wu, Shuang and Bao, Hujun and Huang, Jin},
  journal={IEEE Transactions on Visualization and Computer Graphics}, 
  title={Variational Mesh Offsetting by Smoothed Winding Number}, 
  year={2026},
  volume={32},
  number={2},
  pages={1668-1681},
  doi={10.1109/TVCG.2025.3637845}}

@article{Chen2024,
author = {Chen, Hanyu and Miller, Bailey and Gkioulekas, Ioannis},
title = {3D Reconstruction with Fast Dipole Sums},
year = {2024},
issue_date = {December 2024},
publisher = {Association for Computing Machinery},
address = {New York, NY, USA},
volume = {43},
number = {6},
issn = {0730-0301},
url = {https://doi.org/10.1145/3687914},
doi = {10.1145/3687914},
journal = {ACM Trans. Graph.},
month = nov,
articleno = {192},
numpages = {19}
}

\clearpage

\setcounter{section}{0}
\renewcommand\thesection{\Alph{section}}
\renewcommand\thesubsection{\thesection.\Alph{subsection}}

\section{Detailed Results for the Bandsaw Cutting}

\def\toscaColDist{1cm}

\begin{table}[htbp]
    \sisetup{round-mode=places, round-precision=3, text-series-to-math = true, propagate-math-font = true}
    \centering
    \caption{We compute the average cut volume differences $E_{cut}$ for the TOSCA dataset for each category (both $\times 10^3$). The relative improvement denotes how much unnecessary material of the initial meshes can be removed with the optimization ($1-E_{cut}(\mv{\bar V})/ E_{cut}(\mv V)$). The two-sided Hausdorff distance $\mathbb{H}$ and mean surface distance $\mathbb{M}$ quantify the deformation of the triangle mesh relative to the bounding box diagonal.}
    \label{tab:tosca}

    \begin{filecontents*}{csv/tosca.csv}
Category,ECutOriginal,ECutOptimized,ERelativeImprovement,HDist,MDist
cat,3.4252936179241673,0.45442179014736955,70.77484733648612,0.07455936363636363,0.05178281818181818
centaur,1.721851285158967,0.5318920690721521,67.08620196834563,0.11054183333333334,0.07822016666666667
david,0.7590785805535104,0.35070070589426905,53.01151209449204,0.06702985714285714,0.03599971428571429
dog,2.4343186006363897,0.6025195276985565,71.29895301953616,0.081686,0.059007000000000004
gorilla,56.84639129322022,1.0923768568318337,90.72722956000526,0.14290775,0.11980275
horse,1.6427600203314796,0.4514402025961317,70.89856821057478,0.098992125,0.08195675
michael,3.5082671776763164,0.4934181473799981,57.0500428547432,0.07156789999999999,0.0378823
victoria,4.148036332480842,0.5697933690195592,58.961642167354725,0.10738816666666667,0.056736166666666664
wolf,1.0030782626320918,0.43697250657714903,56.74494962124241,0.09376233333333334,0.06282766666666666
\end{filecontents*}
\csvreader[tabular=lp{\toscaColDist}p{\toscaColDist}p{\toscaColDist}p{\toscaColDist}p{\toscaColDist},
            table head=\toprule
            \bfseries Category & $E_{cut}^{init.}$ & $E_{cut}^{optim.}$ & Improved & $\mathbb{H}$ & $\mathbb{M}$ \\
            \midrule,
            head to column names,
            late after last line=\\\bottomrule]
            {csv/tosca.csv}%
            {}%
            {\Category & \num{\ECutOriginal} & \num{\ECutOptimized} & \num{\ERelativeImprovement}\% & \num{\HDist} & \num{\MDist}}
\end{table}

\section{Pseudocode}

Here we present the full pseudocode for the fast evaluation of the integrated winding numbers given a box filter (\ref{alg:voxelize}) and its derivative in forward-mode (\ref{alg:voxelize_forward}) and backward-mode (\ref{alg:voxelize_backward}).

We used \textsc{IntegrateLinear} as a generalized function for multiple special cases:
Let $\mathbf{1} = (1,\ldots,1)^\top \in \mathbb{R}^d$, $\mathbf{c}^d = (v_0^d, \ldots, v_{d-1}^d)^\top$ (last coords of simplex vertices), $\mathbf{V}_{1:d-1} = \mathbf{P}\mathbf{V}_{\sigma,:,1:d-1}$ (projected Cartesian), $\mathbf{V} = \mathbf{P}\mathbf{V}_\sigma$ (full Cartesian), $\mathbf{I}_d$ ($d$ dimensional identity matrix).
\begin{itemize}
	\item $\textsc{ProjectedArea}(\mathbf{P}, \mathbf{V})$: $\textsc{IntegrateLinear}(\mathbf{P}, \mathbf{V}_{1:d-1}, \mathbf{1})$ \hfill since $\mathbf{b}^\top \mathbf{1} = 1$
	\item $\textsc{AverageHeight}(\mathbf{P}, \mathbf{V}, d)$: $\dfrac{\textsc{IntegrateLinear}(\mathbf{P}, \mathbf{V}_{1:d-1}, \mathbf{c}^d)}{\textsc{IntegrateLinear}(\mathbf{P}, \mathbf{V}_{1:d-1}, \mathbf{1})}$ \hfill since $\mathbf{b}^\top \mathbf{c}^d = x_d$
	\item $\textsc{IntegrateBarycentric}(\mathbf{P}, \mathbf{V})$: $\textsc{IntegrateLinear}(\mathbf{P}, \mathbf{V}, \mathbf{I}_d)$ \hfill (surface measure)
\end{itemize}

\begin{algorithm}
	\caption{Voxelization in $d$ Dimensions}
	\label{alg:voxelize}
	\KwIn{Mesh vertices $V \in \mathbb{R}^{n \times d}$, simplices $S$, grid resolution $\mathbf{r} = (r_1, \ldots, r_d)$}
	\KwOut{Occupancy grid $O \in [0,1]^{r_1 \times \cdots \times r_d}$}
	
	\BlankLine
	$O \gets \mathbf{0}$\tcp*{Initialize occupancy grid}
	$\mathbf{h} \gets (2/r_1, \ldots, 2/r_d)$\tcp*{Voxel size (grid spans $[-1,1]^d$)}
	$m \gets \prod_{\ell=1}^{d-1} h_\ell$\tcp*{Volume of a voxel}
	\BlankLine
	\ForEach{simplex $\sigma \in S$}{
		$\mathbf{v}_0, \ldots, \mathbf{v}_{d-1} \gets$ vertices($\sigma$)\;
		$\mathbf{n} \gets$ normal($\sigma$)\tcp*{Outward pointing normal}
		
		$s \gets \mathrm{sign}(n_d)$\tcp*{Sign of last component of normal}
		
		\BlankLine
		\tcp{Compute bounding box in grid coordinates}
		$(b^1_{\min},\dots,b^d_{\min}), (b^1_{\max},\dots,b^d_{\max}) \gets \textsc{BoundingBox}(\mathbf{V},\mathbf{h})$\;

		\BlankLine
		\tcp{Clip simplex against voxel boundaries}
		$\mathbf{P}^{(0)} \gets \sigma$ in barycentric coordinates\;
		$\mathbf{V}_\sigma \gets (\mathbf{v}_0, \ldots, \mathbf{v}_{d-1})^\top \in \mathbb{R}^{d \times d}$\tcp*{Simplex vertices as matrix}

		\For{$\ell \gets 1$ \KwTo $d$}{
			\For{$i_\ell \gets b^\ell_{\min}$ \KwTo $b^\ell_{\max}$}{
				$t^\ell_{\mathrm{hi}} \gets -1 + (i_\ell + 1) \cdot h_\ell$\;
				$(\mathbf{P}^{(\ell)}, \mathbf{P}^{(\ell)}_{\mathrm{above}}) \gets \textsc{SplitPolytope}(\mathbf{P}^{(\ell-1)}, \mathbf{V}_\sigma, t^\ell_{\mathrm{hi}}, \ell, d)$\;
				$\mathbf{P}^{(\ell-1)} \gets \mathbf{P}^{(\ell)}_{\mathrm{above}}$\tcp*{Remaining for next iteration}

				\BlankLine

				\If{$\ell = d$}{
					\tcp{Compute Cartesian coords of clipped polytope}
					$\mathbf{V}^{(d)} \gets \mathbf{P}^{(d)} \mathbf{V}_\sigma \in \mathbb{R}^{k \times d}$\;
					\BlankLine
					\tcp{Contribution at voxel $(i_1, \ldots, i_d)$:}
					$a \gets \textsc{ProjectedArea}(\mathbf{P}^{(d)}, \mathbf{V}^{(d)})$\tcp*{onto first $d-1$ axes (footprint)}
                    \BlankLine
					\For{$k \gets i_d-1$ \KwTo $0$}{
						\tcp{For all voxels below the current}
						$\mathbf{i_k} \gets (i_1, \ldots, i_{d-1},k)$\;
						$O[\mathbf{i_k}] \gets O[\mathbf{i_k}] + s \cdot a \cdot h_d / m$\;
					}
					\BlankLine
					$\mathbf{i} \gets (i_1, \ldots, i_d)$\;
					$v^{\text{avg}}_d \gets \textsc{AverageHeight}(\mathbf{P}^{(d)}, \mathbf{V}^{(d)}, d)$\tcp*{in $d$ direction}
					$O[\mathbf{i}] \gets O[\mathbf{i}] + s \cdot a \cdot (v^{avg}_d - (t^d_{\mathrm{hi}} - h_d))/ m$\;
				}
			}
		}
	}
\end{algorithm}

\begin{algorithm}
\caption{Forward-Mode Derivatives for Voxelization}
\label{alg:voxelize_forward}
\KwIn{Mesh vertices $V \in \mathbb{R}^{n \times d}$, simplices $S$, grid resolution $\mathbf{r} = (r_1, \ldots, r_d)$, vertex tangents $\dot{V} \in \mathbb{R}^{n \times d}$}
\KwOut{Occupancy tangent $\dot{O} \in \mathbb{R}^{r_1 \times \cdots \times r_d}$}

\BlankLine
$\mathbf{h} \gets (2/r_1, \ldots, 2/r_d)$\tcp*{Voxel size (grid spans $[-1,1]^d$)}
$m \gets \prod_{\ell=1}^{d} h_\ell$\tcp*{Volume of a voxel}
\BlankLine
\ForEach{simplex $\sigma \in S$}{
	$\mathbf{v}_0, \ldots, \mathbf{v}_{d-1} \gets$ vertices($\sigma$)\;
	$\hat{\mathbf{n}} \gets$ unit\_normal($\sigma$)\tcp*{Unit outward normal}

	\BlankLine
	\tcp{Project vertex tangents onto normal direction}
	\For{$i \gets 0$ \KwTo $d-1$}{
		$\dot{v}^n_i \gets \dot{\mathbf{v}}_i \cdot \hat{\mathbf{n}}$\tcp*{Normal component of tangent}
	}

	\BlankLine
    
    \tcp{Compute bounding box (same as primal)}
    $(b^1_{\min},\dots,b^d_{\min}), (b^1_{\max},\dots,b^d_{\max}) \gets \textsc{BoundingBox}(\mathbf{V},\mathbf{h})$\;

	\BlankLine
	\tcp{Clip simplex against voxel boundaries (same as primal)}
	$\mathbf{P}^{(0)} \gets \sigma$ in barycentric coordinates\;
	$\mathbf{V}_\sigma \gets (\mathbf{v}_0, \ldots, \mathbf{v}_{d-1})^\top \in \mathbb{R}^{d \times d}$\;

	\For{$\ell \gets 1$ \KwTo $d$}{
		\For{$i_\ell \gets b^\ell_{\min}$ \KwTo $b^\ell_{\max}$}{
			$t^\ell_{\mathrm{hi}} \gets -1 + (i_\ell + 1) \cdot h_\ell$\;
			$(\mathbf{P}^{(\ell)}, \mathbf{P}^{(\ell)}_{\mathrm{above}}) \gets \textsc{SplitPolytope}(\mathbf{P}^{(\ell-1)}, \mathbf{V}_\sigma, t^\ell_{\mathrm{hi}}, \ell, d)$\;
			$\mathbf{P}^{(\ell-1)} \gets \mathbf{P}^{(\ell)}_{\mathrm{above}}$\;

			\BlankLine

			\If{$\ell = d$}{
				$\mathbf{i} \gets (i_1, \ldots, i_d)$\;
				$\mathbf{V}^{(d)} \gets \mathbf{P}^{(d)} \mathbf{V}_\sigma$\tcp*{Cartesian coords of clipped polytope}
				\BlankLine
				\tcp{Integrate normal velocity over clipped polygon (surface area)}
				$\dot{O}[\mathbf{i}] \gets \dot{O}[\mathbf{i}] + \textsc{IntegrateLinear}(\mathbf{P}^{(d)}, \mathbf{V}^{(d)}, \dot{\mathbf{v}}^n) / m$\;
			}
		}
	}
}
\end{algorithm}

\begin{algorithm}
\caption{Backward-Mode Derivatives for Voxelization}
\label{alg:voxelize_backward}
\KwIn{Mesh vertices $V \in \mathbb{R}^{n \times d}$, simplices $S$, grid resolution $\mathbf{r} = (r_1, \ldots, r_d)$, occupancy adjoint $\bar{O} \in \mathbb{R}^{r_1 \times \cdots \times r_d}$}
\KwOut{Vertex adjoint $\bar{V} \in \mathbb{R}^{n \times d}$}

\BlankLine
$\mathbf{h} \gets (2/r_1, \ldots, 2/r_d)$\tcp*{Voxel size (grid spans $[-1,1]^d$)}
$m \gets \prod_{\ell=1}^{d} h_\ell$\tcp*{Volume of a voxel}
\BlankLine
\ForEach{simplex $\sigma \in S$}{
	$\mathbf{v}_0, \ldots, \mathbf{v}_{d-1} \gets$ vertices($\sigma$)\;
	$\hat{\mathbf{n}} \gets$ unit\_normal($\sigma$)\tcp*{Unit outward normal}

	\BlankLine
    \tcp{Compute bounding box (same as primal)}
    $(b^1_{\min},\dots,b^d_{\min}), (b^1_{\max},\dots,b^d_{\max}) \gets \textsc{BoundingBox}(\mathbf{V},\mathbf{h})$\;

	\BlankLine
	\tcp{Clip simplex against voxel boundaries (same as primal)}
	$\mathbf{P}^{(0)} \gets \sigma$ in barycentric coordinates\;
	$\mathbf{V}_\sigma \gets (\mathbf{v}_0, \ldots, \mathbf{v}_{d-1})^\top \in \mathbb{R}^{d \times d}$\;

	\For{$\ell \gets 1$ \KwTo $d$}{
		\For{$i_\ell \gets b^\ell_{\min}$ \KwTo $b^\ell_{\max}$}{
			$t^\ell_{\mathrm{hi}} \gets -1 + (i_\ell + 1) \cdot h_\ell$\;
			$(\mathbf{P}^{(\ell)}, \mathbf{P}^{(\ell)}_{\mathrm{above}}) \gets \textsc{SplitPolytope}(\mathbf{P}^{(\ell-1)}, \mathbf{V}_\sigma, t^\ell_{\mathrm{hi}}, \ell, d)$\;
			$\mathbf{P}^{(\ell-1)} \gets \mathbf{P}^{(\ell)}_{\mathrm{above}}$\;

			\BlankLine

			\If{$\ell = d$}{
				$\mathbf{i} \gets (i_1, \ldots, i_d)$\;
				$w \gets \bar{O}[\mathbf{i}] / m$\tcp*{Adjoint weight}
				$\mathbf{V}^{(d)} \gets \mathbf{P}^{(d)} \mathbf{V}_\sigma$\tcp*{Cartesian coords of clipped polytope}
				\BlankLine
				\tcp{Distribute adjoint to simplex vertices via barycentric coords (surface area)}
				$(\bar{b}_0, \ldots, \bar{b}_{d-1}) \gets \textsc{IntegrateBarycentric}(\mathbf{P}^{(d)}, \mathbf{V}^{(d)})$\;
				\BlankLine
				\tcp{Accumulate to vertex adjoints in normal direction}
				\For{$j \gets 0$ \KwTo $d-1$}{
					$\bar{\mathbf{v}}_j \gets \bar{\mathbf{v}}_j + w \cdot \bar{b}_j \cdot \hat{\mathbf{n}}$\;
				}
			}
		}
	}
}
\end{algorithm}

\begin{algorithm}
\caption{\textsc{SplitPolytope}: Split a polytope by a hyperplane}
\label{alg:split_polytope}
\KwIn{Polytope $\mathbf{P}$ in barycentric coords, simplex vertices $\mathbf{V}_\sigma \in \mathbb{R}^{d \times d}$, threshold $t$, axis $\ell$, dimension $d$}
\KwOut{$(\mathbf{P}_{\mathrm{below}}, \mathbf{P}_{\mathrm{above}})$: polytopes below/above the hyperplane $x_\ell = t$}

\BlankLine
\eIf{$d = 2$}{
	\Return{\textsc{SplitSegment}$(\mathbf{P}, \mathbf{V}_\sigma, t, \ell)$}\;
}{
	\Return{\textsc{SplitPolygon}$(\mathbf{P}, \mathbf{V}_\sigma, t, \ell)$}\;
}
\end{algorithm}

\begin{algorithm}
\caption{\textsc{SplitSegment}: Split a segment by a line ($d=2$)}
\label{alg:split_segment}
\KwIn{Segment $\mathbf{P} = (t_{\mathrm{lo}}, t_{\mathrm{hi}})$ as parameter range, edge vertices $\mathbf{V}_\sigma \in \mathbb{R}^{2 \times 2}$, threshold $t$, axis $\ell$}
\KwOut{$(\mathbf{P}_{\mathrm{below}}, \mathbf{P}_{\mathrm{above}})$: segments below/above the line $x_\ell = t$}

\BlankLine
\tcp{Cartesian coords at segment endpoints}
$x_{\mathrm{lo}} \gets (1 - t_{\mathrm{lo}}) \cdot V_\sigma^{0,\ell} + t_{\mathrm{lo}} \cdot V_\sigma^{1,\ell}$\;
$x_{\mathrm{hi}} \gets (1 - t_{\mathrm{hi}}) \cdot V_\sigma^{0,\ell} + t_{\mathrm{hi}} \cdot V_\sigma^{1,\ell}$\;
\BlankLine
$b_{\mathrm{lo}} \gets (x_{\mathrm{lo}} \le t)$; $b_{\mathrm{hi}} \gets (x_{\mathrm{hi}} \le t)$\;
\BlankLine
\uIf{$b_{\mathrm{lo}} \land b_{\mathrm{hi}}$}{
	\tcp{Entire segment below}
	\Return{$((t_{\mathrm{lo}}, t_{\mathrm{hi}}), \emptyset)$}\;
}
\uElseIf{$\neg b_{\mathrm{lo}} \land \neg b_{\mathrm{hi}}$}{
	\tcp{Entire segment above}
	\Return{$(\emptyset, (t_{\mathrm{lo}}, t_{\mathrm{hi}}))$}\;
}
\Else{
	\tcp{Segment crosses the line}
	$t_{\mathrm{mid}} \gets t_{\mathrm{lo}} + (t - x_{\mathrm{lo}}) / (x_{\mathrm{hi}} - x_{\mathrm{lo}}) \cdot (t_{\mathrm{hi}} - t_{\mathrm{lo}})$\;
	\eIf{$b_{\mathrm{lo}}$}{
		\Return{$((t_{\mathrm{lo}}, t_{\mathrm{mid}}), (t_{\mathrm{mid}}, t_{\mathrm{hi}}))$}\;
	}{
		\Return{$((t_{\mathrm{mid}}, t_{\mathrm{hi}}), (t_{\mathrm{lo}}, t_{\mathrm{mid}}))$}\;
	}
}
\end{algorithm}

\begin{algorithm}
\caption{\textsc{SplitPolygon}: Split a polygon by a plane ($d=3$)}
\label{alg:split_polygon}
\KwIn{Polygon $\mathbf{P}$ in barycentric coords, triangle vertices $\mathbf{V}_\sigma \in \mathbb{R}^{3 \times 3}$, threshold $t$, axis $\ell$}
\KwOut{$(\mathbf{P}_{\mathrm{below}}, \mathbf{P}_{\mathrm{above}})$: polygons below/above the plane $x_\ell = t$}

\BlankLine
$\mathbf{P}_{\mathrm{below}} \gets []$; $\mathbf{P}_{\mathrm{above}} \gets []$\;
$\mathbf{c}^\ell \gets \mathbf{V}_{\sigma,:,\ell}$\tcp*{$\ell$-th coord of simplex vertices}
\BlankLine
\tcp{Sutherland-Hodgman style clipping}
\For{$i \gets 0$ \KwTo $|\mathbf{P}|-1$}{
	$\mathbf{b}_{\mathrm{curr}} \gets \mathbf{P}[i]$; $\mathbf{b}_{\mathrm{next}} \gets \mathbf{P}[(i+1) \mod |\mathbf{P}|]$\;
	$x_{\mathrm{curr}} \gets \mathbf{b}_{\mathrm{curr}} \cdot \mathbf{c}^\ell$\tcp*{Cartesian coord via barycentric interp}
	$x_{\mathrm{next}} \gets \mathbf{b}_{\mathrm{next}} \cdot \mathbf{c}^\ell$\;
	\BlankLine
	\tcp{Classify current vertex}
	\eIf{$x_{\mathrm{curr}} \le t$}{
		$\mathbf{P}_{\mathrm{below}}$.append($\mathbf{b}_{\mathrm{curr}}$)\;
	}{
		$\mathbf{P}_{\mathrm{above}}$.append($\mathbf{b}_{\mathrm{curr}}$)\;
	}
	\BlankLine
	\tcp{Check for edge crossing}
	\If{$(x_{\mathrm{curr}} \le t) \neq (x_{\mathrm{next}} \le t)$}{
		$\alpha \gets (t - x_{\mathrm{curr}}) / (x_{\mathrm{next}} - x_{\mathrm{curr}})$\;
		$\mathbf{b}_{\mathrm{int}} \gets (1-\alpha) \cdot \mathbf{b}_{\mathrm{curr}} + \alpha \cdot \mathbf{b}_{\mathrm{next}}$\tcp*{Intersection point}
		$\mathbf{P}_{\mathrm{below}}$.append($\mathbf{b}_{\mathrm{int}}$)\;
		$\mathbf{P}_{\mathrm{above}}$.append($\mathbf{b}_{\mathrm{int}}$)\;
	}
}
\Return{$(\mathbf{P}_{\mathrm{below}}, \mathbf{P}_{\mathrm{above}})$}
\end{algorithm}

\begin{algorithm}
\caption{\textsc{IntegrateLinear}: Integrate linear function(s) over clipped polytope}
\label{alg:integrate_linear}
\KwIn{Polytope $\mathbf{P}\in \mathbb{R}^{k \times d}$ (barycentric), $\mathbf{V} \in \mathbb{R}^{k \times n}$ (Cartesian), $\mathbf{F} \in \mathbb{R}^{d \times m}$ (values)}
\KwOut{$\mathbf{I} \in \mathbb{R}^m$ where $I_j = \int_P f_j(\mathbf{x}) \, d\mu$ and $f_j(\mathbf{b}) = \mathbf{b}^\top \mathbf{F}_{:,j}$}

\BlankLine
$\mathbf{I} \gets \mathbf{0} \in \mathbb{R}^m$\;
\BlankLine
\tcp{Triangulate from first vertex into $(d{-}1)$-simplices}
\For{$i \gets 1$ \KwTo $|\mathbf{P}|-(d-1)$}{
	\tcp{Sub-simplex measure (length for $d{=}2$, area for $d{=}3$)}
	$a_i \gets \textsc{Measure}_{d-1}(\mathbf{V}_{0}, \mathbf{V}_{i}, \ldots, \mathbf{V}_{i+d-2})$\;
	\BlankLine
	\tcp{Average barycentric coords over sub-simplex}
	$\bar{\mathbf{b}}_i \gets (\mathbf{P}_0 + \mathbf{P}_i + \cdots + \mathbf{P}_{i+d-2}) / d$\;
	\BlankLine
	\tcp{Accumulate weighted function values}
	$\mathbf{I} \gets \mathbf{I} + a_i \cdot \bar{\mathbf{b}}_i^\top \mathbf{F}$\;
}
\Return{$\mathbf{I}$}
\end{algorithm}

\begin{algorithm}
\caption{\textsc{BoundingBox}: Compute bounding box of simplex in grid coordinates}
\label{alg:integrate_linear}
\KwIn{Simplex $\mathbf{V} \in \mathbb{R}^{d \times n}$ (Cartesian), Voxel size $\mathbf{h}\in\mathbb{R}^d$}
\KwOut{minimal voxel in each dimension $\mathbf{b}_{\min} = (b^1_{\min},\dots,b^d_{\min})$, maximal voxel in each dimension $\mathbf{b}_{\max} = (b^1_{\max},\dots,b^d_{\max})$}

\BlankLine
\tcp{Compute bounding box in grid coordinates}
\For{$\ell \gets 1$ \KwTo $d$}{
    $b^\ell_{\min} \gets \max\left(0, \left\lfloor \frac{\min_i v_i^\ell + 1}{h_\ell} \right\rfloor\right)$\;
    $b^\ell_{\max} \gets \min\left(r_\ell - 1, \left\lfloor \frac{\max_i v_i^\ell + 1}{h_\ell} \right\rfloor\right)$\;
}
\end{algorithm}

\clearpage

\end{document}